\documentclass[preprint,aps,showpacs]{revtex4}
\usepackage{color}
\usepackage{colordvi}
\usepackage{amssymb}
\usepackage{amsmath}
\usepackage{epsf}
\usepackage{mathrsfs}
\usepackage{graphicx}
\usepackage{appendix}

\usepackage{subfig}

\begin{document}
\def \beq{\begin{equation}}
\def \eeq{\end{equation}}
\def \bea{\begin{eqnarray}}
\def \eea{\end{eqnarray}}
\def \bem{\begin{displaymath}}
\def \eem{\end{displaymath}}
\def \P{\Psi}
\def \Pd{|\Psi(\boldsymbol{r})|}
\def \Pds{|\Psi^{\ast}(\boldsymbol{r})|}
\def \Po{\overline{\Psi}}
\def \bs{\boldsymbol}
\def \bl{\bar{\boldsymbol{l}}}
\title{ Hydrodynamic theory of Rotating Ultracold Bose Einstein Condensates in Supersolid Phase}
\author{ Rashi Sachdeva \footnote{Present address: Quantum systems unit, Okinawa Institute of Science and Technology, Okinawa, Japan}, Sankalpa Ghosh }
\affiliation{Department of Physics, Indian Institute of Technology, Delhi, New Delhi-110016}
\begin{abstract}
Within mean field Gross-Pitaevskii framework, ultra cold atomic condensates with long range interaction is predicted to have a supersolid like ground state beyond a critical  interaction strength. Such  mean field supersolid like ground state has periodically modulated superfluid density which implies the coexistence of superfluid and crystalline order.  Ultra cold atomic system in such mean field ground state can be subjected to artificial gauge field created either through rotation or
by introducing space dependent coupling among hyperfine states of the atoms using Raman lasers.
Starting from this  Gross-Pitaevskii energy functional that describes such systems at zero temperature, we construct 
hydrodynamic theory to describe the low energy long wavelength excitations of such 
 rotating supersolid of weakly interacting ultra cold atoms in two spatial dimensions
for generic type of long range interaction. We treat the supersolidity in such system within the framework of
well known two fluid approximation. Considering such  system
in the fast rotation limit where a  vortex lattice in superfluid coexists  with the supersolid lattice, we analytically obtain the dispersion relations of collective excitations around this equilibrium state. The dispersion relation gives the modes of the rotating supersolid which can be experimentally measured within the current technology. We point out that this can clearly identify such a ultra cold atomic supersolid phase in an unambiguous way.
\end{abstract}
\pacs{03.75.-b, 67.85.-d, 67.80.bd}
\date{\today}
\maketitle
\section{Introduction}
The issue of observing supersolidity experimentally in solid $^{4} He$ \cite{Mosechan, Reppy}  has now settled conclusively  
by showing that there is no supersolidity in such system \cite{Mosechan1}. As a result, the counter-intuitive co-existence of superfluidity and crystalline order
\cite{PO,Andreev, Legget, ODLRO, Anderson} still remains an open question inspite of lots of progress in this direction \cite{review}. In this context, an alternative possible route to observe supersolidity in much more controllable and conspiciuous way is via certain species of ultra cold atomic condensate with long range interaction. These ultracold atom condensates with long range interactions can have roton-instability in their excitation spectrum \cite{santos, Pomeau, Nozieres} and significant experimental  \cite{dipole1, Rydberg, 
polar1, Grimm, Lev} as well as theoretical \cite{He, Jain, Rejish, Cinti, Li}  progress took place 
in realizing such systems. In recent experiments  such roton like mode softening has been demonstrated through cavity mediated long-range interaction in ultra cold atomic BEC \cite{RotonEx} and a self organized supersolid phase has also been experimentally observed \cite{expt2} in Dicke quantum phase transition where the long-range interaction is generated by a two-photon process in cavity. 

In this work, we show that one way of clearly identifying such ultra cold supersolid phase is to study 
its response to an artificial gauge field created through rotation or by other means \cite{rotation, rotlat, synthetic}. 
Study of the critical velocity of nucleation of vortices in a rotating dipole-blockaded ultracold supersolid condensate \cite{Mason}
as well as supersolid vortex lattice phases in a fast rotating Rydberg dressed Bose-Einstein condensate within Gross-Pitaevskii approach
\cite{ssvortex} were carried out recently. The same work \cite{ssvortex} particularly brought forward important difference in the vortex lattice structure in such supersolid like ground state 
as compared to similar vortex lattice structure in ultra cold atomic superfluid state. However,
it is still not clear if within the standard time of flight measurement technique, one will be able to separately identify the vortex cores in vortex lattices from the superfluid density 
minimum in the supersolid lattices. A way out from this problem is to look for the collective excitation spectrum 
of such supersolid vortex lattices.

In an ultracold atomic ensemble with long range interaction, a supersolid like ground state implies a periodic modulation of the superfluid density when the relative strength of such interaction exceeds a critical value. This implies that the supersolid phase possess phase coherence as well as periodic density distribution, which results in density modulated superfluid, where the density maxima or minima forms a lattice, referred to as supersolid lattice in this paper. It is to note that this is completely different from the density wave phase, which is an insulating phase with no phase coherence, but possess a periodic or crystalline distribution of particles, with no superfluidity. Typically for weakly interacting bosons near absolute zero temperature, such an ultra cold Bose Einstein Condensate is theoretically 
described within the framework of Gross Pitaevskii equation for short range as well as for generic long range interaction in mean field approximation.  
Within this framework a periodic modulation was discussed as early as in 1957 by E. Gross \cite{Gross} and recently discussed in several contexts \cite{Pomeau, Rejish, review} that include ultra cold atomic systems.  
Such a supersolid ground state is different from the Andreev-Lifshitz supersolid scenario \cite{Andreev} which is based on vacancies or interstitials with 
repulsive interactions, more appropriate for the solid $^4$He. 

\begin{figure}[htb]
\begin{center}
{ \includegraphics[width=0.9\textwidth]{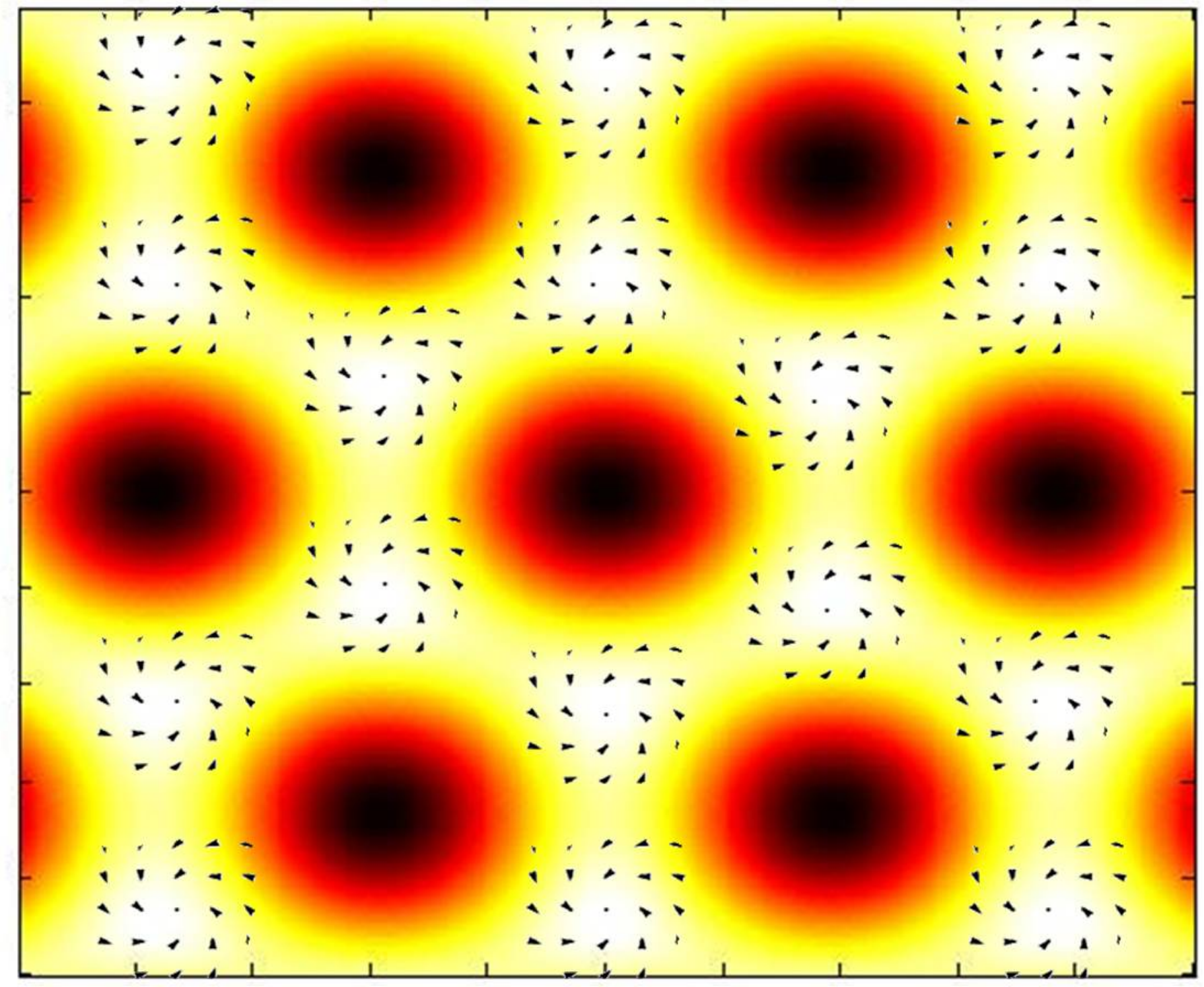}}
\caption{Schematic figure showing the co-existing vortex lattice and supersolid lattice in a rotating supersolid. The high density (dark) areas show the supersolid crystal which is hexagonal in shape, and the low density (light) areas show the vortex lattice, with arrow plots showing the winding of single vortex. }

\label{vlss}
\end{center}
\end{figure}

We study the effect of sufficiently high artificial magnetic field on such a supersolid phase which results in formation of vortex lattice phase in such system. As shown in recent literature \cite{ssvortex} such vortices can arrange themselves either at the minima or maxima of the supersolid density periodic modulation. 
In more specific terms, the superfluid density is modulated in a periodic manner in the supersolid phase.
When such supersolid is rotated fast a vortex lattice is formed and there is modulation of superfluid density 
due to the formation of vortices. Particularly at the core of such vortices the superfluid density goes 
to zero and there will be circulation around such vortex core. This vortex core may coincide with the minima 
as well as the maxima of the superfluid density in the superfluid phase ( ref. \cite{ssvortex}) under various conditions. For 
example in Fig. \ref{vlss} we schematically described a situation where the vortex lattice co-exists with supersolid lattice and the 
vortex core coincides with the superfluid density minimum in the supersolid phase. 

The high density (dark) areas in Fig. \ref{vlss} show the supersolid crystal lattice (hexagonal), whereas the low density (light) areas show the vortex positions superposed with arrow plots to show the winding of single vortex. Treating the small oscillation around such equilibrium state in the low energy long wavelength limit,  we construct a hydrodynamic theory of collective excitations of such a  vortex lattice state in ultra cold atomic supersolid.  Particularly we calculate the  dispersion of such low energy long wavelength collective excitations and explain how they demonstrate the supersolid behavior.

In this paper, in the framework of a hydrodynamic theory,
we demonstrate 
that the collective excitation spectrum of vortex lattice phase in a fast rotating ultra cold atomic 
supersolid  have important 
differences with the collective excitation of the Abrikosov vortex lattice phase in ultra cold atomic superfluid. A lot of studies have been done for rapidly rotating BEC with their vortex lattices and their collective excitations \cite{Coddington, tkachenko, mizushima, Baym3, Sonin1, Baym, Baym1, Sonin, Fetterreview}. The collective excitations  have been studied within hydrodynamic framework \cite{ Sonin, Baym, Baym1} where Tkachenko modes \cite{Coddington} have been studied and compared with similar theories earlier developed for superfluid Helium \cite{Baym3, Sonin1}.  We argue that experimental detection of the same for a rotating supersolid can provide a conclusive test of supersolidity, and hence motivates the present work. 
The adaptation of  hydrodynamic theory in the present case  implies writing the equations of motion in terms of density and phase, and describing  the long wavelength behavior of the fluid with these equations.  To validate such hydrodynamic approximation, the variables used in these equations are averaged over scales much larger than the inter vortex spacing and the supersolid lattice spacing. 

For low $q$ values, we show that our analytical results also qualitatively agree with the appearance of two distinct longitudinal modes for a supersolid, in a recent work by Saccani et.al \cite{boninsegni} derived from a microscopic model using quantum Monte carlo method. Our results show the appearance of longitudinal as well as transverse modes of rotating supersolid by analytical means, and can be confirmed by numerical calculations qualitatively. A quantitative comparison of the results requires much more involved analytical calculations, including the mutual friction of the two co-existing lattices and also the different possible symmetry considerations of the vortex and the supersolid lattice, which is out of context of the present work.

The rest of the paper is organized in the following way. In section \ref{HydLag} we derived the hydrodynamic Lagrangian from the Gross Pitaevskii Energy 
functional using homogenization method. Section {\ref{HydEq} shows the determination of hydrodynamic equations of motion, the calculation of dispersion relations and the corresponding sound modes for the rotating ultracold supersolid. We conclude by emphasizing the significance of the main finding of this work, namely the dispersion modes for the rotating supersolid system, and also point out the possibility of experimental verification. The other details of the calculations are provided in Appendix $A$.

\section{Effective Hydrodynamic Lagrangian}\label{HydLag}
We begin with a Gross-Pitaevskii mean field description of ultra cold atomic Bose Einstein condensate at $T=0$ with suitable long range interaction, rotated about the $z$-axis with a frequency $\Omega$  in two dimensions.  It may be pointed out here that our equilibriuim state is the one obtained in the limit of high rotation, the trap potential is almost cancelled by the centrifugal force \cite{newref} and the system can be very well approximated as a uniform  two dimensional system. The details of the derivation of such Gross-Pitaevskii energy functional from the microscopic 
Hamiltonian of a  typical ultra cold system such as Rydberg excited BEC is given in \cite{Rejish}.  
As already mentioned, at a sufficiently large interaction strength and
fast enough rotational frequency $\Omega$,  the ground state of the system is a vortex lattice  phase of  the supersolid,
as shown by a recent numerical study \cite{ssvortex}. We are interested in low energy 
excitations of such system which has wavelength  much larger than the lattice parameters of the vortex lattice or the supersolid lattice. 
 
The mean field  Lagrangian for the rotating supersolid system is given as
\beq  \textit{L} = 
\int d \mathbf{r}\left[\frac{i\hbar}{2} \left(\psi^{*}\frac{\partial \psi}{\partial t} -\psi \frac{\partial \psi^{*}}{\partial t}\right) - \mathcal{E}_{R}(\psi,\psi^{*}) \right], \label{Lag} \eeq 
Here $\mathcal{E}_{R}$ is the  Gross Pitaevskii energy functional, in the co-rotating frame \cite{Book1, Book2}, related with the non rotating energy functional $\mathcal{E}$ through 
the expression $ \mathcal{E}_{R} = \mathcal{E} -  \Omega \cdot <( \mathbf{r} \times \mathbf{p})>$. $\mathcal{E}$ is   the 
usual Gross-Pitaevskii energy functional,  given by
\beq \mathcal {E} (\psi, \psi^{*})= \frac{\hbar^2}{2m} |\nabla \psi|^2  + \frac{1}{2}
\int d \mathbf{r'}|\psi(\mathbf{r})|^2U(|\mathbf{r}-\mathbf{r}'|) |\psi(\mathbf{r}')|^2 \label{gpenergyfunc}\eeq

In the usual zero temperature mean field description of an ultracold atomic superfluid \cite{Book1}, $|\psi(\bs{r})|^{2}$ is identified with superfluid density and $\psi(\bs{r})$ as the superfluid order parameter. 
However in the present case for an ultracold atomic supersolid \cite{Rica2, newrica,Rica3}, one can extract a Landau two fluid description from the 
same Gross Pitaevskii energy functional, where the normal component of the two fluid description 
corresponds to the solid part of the supersolid. We must mention at the outset that from now onwards, the superscript 'ss' stands for the supersolid lattice component, which plays the role of the normal component in the two-fluid desciption and superscript 'v' stands for the vortex lattice component of the system, in our subsequent calculations. 

To do this we first write complex   $\psi(\mathbf{r},t)=\sqrt{n(\mathbf{r},t)}e^{i\Phi(\mathbf{r},t)}$ in terms of the density $n$ and phase $\Phi$. Then Lagrangian $\textit{L}$ in the 
non-rotating case  takes the form
\beq \textit{L}  =\int  d\bs{r} \left\{-\left[\hbar n \frac{\partial \Phi}{\partial t} + \frac{\hbar^{2}}{2m}\left(n(\nabla\Phi)^{2}+\frac{1}{4 n}(\nabla n)^{2}\right)
\right]d\mathbf{r}-\frac{1}{2}\int U(|\mathbf{r}-\mathbf{r'}|)n(\mathbf{r})n(\mathbf{r'})d\mathbf{r'}\right\}\label{lagrhophi}\eeq

We also introduce $\mathbf{u}^{ss}(\mathbf{r},t)$ as the displacement field of the supersolid lattice.
In an ordinary superfluid the average superfluid density 
$\rho=\frac{1}{V}\int_{V}n(\mathbf{r})d\mathbf{r}$ is constant, where $V$ is the total volume of the system.
On the other hand, in a given classical crystalline solid, $\rho$  is defined by a  fixed number of atoms per unit cell
for a given set of lattice vectors  such that the elastic deformation of lattice 
parameters obeys $\frac{\delta \rho}{\rho}=-\nabla\cdot \mathbf{u}^{ss}$. 
In an ultra cold atomic supersolid, quantum fluctuation leads to additional compression/dilation effects of the lattice ($\mathbf{u}^{ss}(\mathbf{r},t$)) which adds to the superfluid density.
This is basically due to the change in
 displacement field $\mathbf{u}^{ss}(\mathbf{r},t$)  or equivalently by changing the density of superfluid component. Hence, this fact can be expressed through the following ansatz \cite{Rica3} 
\beq \rho=\bar{\rho}+\bar{\rho}\nabla\cdot\mathbf{u^{ss}}+\delta\rho \label{rhoexpand}\eeq
The above relation takes into account the fluctuations in the density of the supersolid.
It is to note that $\rho$ is the total density, which comprises of the superfluid density and the crystal density due to spatial modulations in density. We describe the lattice part of the supersolid as the normal component within the well known two fluid description.  $\delta\rho$ are the fluctuations around the steady state with density $\bar{\rho}$. For a usual prototype superfluid, $\rho$ is simply the superfluid density with $\delta\rho$ as the fluctuations around the steady state.

In the same way as in a crystalline solid where the presence of a lattice makes the effective mass of electron as a tensor, here also in the presence of a lattice like normal component, the superfluid density will be tensor like quantity \cite{Paananen} in a supersolid.  For typical lattice structure such as hcp and fcc lattice, it has been shown \cite{isotropicss} that the superfluid flow is same in all directions of the crystal and hence one can write the  superfluid density tensor in an isotropic form. In the current work  we also  consider an isotropic supersolid so that the structure of this tensor is purely diagonal and is given by $ \rho_{ik}^{ss}=\rho^{ss}(\rho)\delta_{ik}$ with all components having the same value.

When the system is rotated fast enough, a vortex lattice is formed that can also be characterized as a patterned modulation in the superfluid density and phase. 
To denote the fluctuations of the vortex lattice from its equilibrium position, we introduce displacement field  $\mathbf{u}^{v}(\mathbf{r},t)$.
This vortex lattice has an associated vortex crystal lattice effective mass.
For the case of rotating superfluids such an effective mass for the vortex lattice was considered in the literature 
\cite{Baym3}. When such effective  mass is taken into account, it leads to an additional term in the kinetic 
energy of the system  which will be proportional to the product of mass density of vortex lattice and square of velocity difference between the superfluid and vortex lattice velocity. Additionally in the present case, it will
also produce terms due to the relative motion between the vortex lattice and the normal component due to the supersolid lattice. This fact can be appreciated also by  inspecting the expression of  the Lagrangian (\ref{Lageff}) and $\mathcal{E}_{ph}(\phi)$ in the subsequent derivation.
In a less technical language by 
introduction of vortex lattice effective mass, the system will have mutual friction or relative motion between the different components. To simplify further analysis, we ignore such relative motions that arise due to the effective mass of the vortex lattice, and take into account only the supersolid crystal effective mass. This approximation has also been explained and shown in detail mathematically in appendix \ref{app11}.

Thus the displacement field $\mathbf{u}(\mathbf{r},t)$ as well as the average density $\rho$ can be varied independently
and hence, the complex macroscopic wavefunction $\psi(\mathbf{r},t)$ is now a functional of
three field variables $\rho(\mathbf{r},t),\mathbf{u}(\mathbf{u}^{ss}(\mathbf{r},t), \mathbf{u}^{v}(\mathbf{r},t))$ and $\phi(\mathbf{r},t)$.
To construct a long wavelength description of the system we use the homogenization technique \cite{Rica2,newrica, Rica3}
 in which one separates the density
and phase in fast and slow varying components, and the fast varying component is integrated out.
This finally gives us the effective Lagrangian as 
\beq  L =  \int d \mathbf{r} \left[-\hbar \rho \frac{\partial \phi}{\partial t} - \mathcal{E}\right]  \label{Lageff} \eeq
where
\bea \mathcal{E} &= & \mathcal{E}_{in}(\rho)+\mathcal{E}_{ph}(\phi)+\mathcal{E}_{el}^{ss}(\nabla\mathbf{u}^{ss})+\mathcal{E}_{el}^{v}(\nabla\mathbf{u}^{v})\label{totalsum}   \\
\mathcal{E}_{in}(\rho)&=& \frac{\hbar^{2}}{2m}\frac{(\nabla\rho)^{2}}{4\rho}+\rho\int U(|\mathbf{r}-\mathbf{r}'|)\rho(\mathbf{r}')d\mathbf{r}'=\mu\rho\label{internal}
\\
\mathcal{E}_{ph}(\phi)&=&\frac{\hbar^{2}}{2m}[\rho(\nabla\phi)^{2}-(\rho\delta_{ik}-\rho_{ik}^{ss})
\left(\nabla\phi-\frac{m}{\hbar}\frac{D\mathbf{u}^{ss}}{Dt}\right)_{i} \nonumber \\
&  & \left(\nabla\phi-\frac{m}{\hbar}\frac{D\mathbf{u}^{ss}}{Dt}\right)_{k}]
 +m\rho\mathbf{v}_{s}\cdot(\Omega\times\mathbf{r})\label{phasecont}\\
\mathcal{E}_{el}^{ss}(\nabla\mathbf{u}^{ss})&=& \frac{1}{2}\lambda_{iklm}^{s}\epsilon_{ik}^{s}\epsilon_{lm}^{s}
;~~ \mathcal{E}_{el}^{v}(\nabla\mathbf{u}^{v})= \frac{1}{2}\lambda_{iklm}^{v}\epsilon_{ik}^{v}\epsilon_{lm}^{v}\label{allpartsnew}\eea
The details of the derivation \cite{supp1} is given in Appendix \ref{app11} (section 1).

 Let us briefly summarize the main approximations that we have made to arrive at the effective energy functional. As mentioned earlier, we consider a rapidly rotating condensate where the rotation takes place in the $x-y$ plane about the $z$-axis with rotation frequency $\Omega$ being very close to
the two dimensional trapping potential $\omega_{\perp}$ \cite{newref}.  
Therefore, the effective trapping  potential in the $x-y$ plane  is given by  $V_{ext}=\frac{1}{2}m (\omega_{\perp}^{2} - \Omega^{2})r^2$  and for such fast rotating condensate, it is set to zero for the rest of the calculation.  In this regard we may point out that in experiments on rotating ultra cold Bose Einstein condensates the rotational frequency frequency $\Omega$ as high as  $0.99$ 
part of the transverse trapping frequency $\omega_{\perp}$ was achieved \cite{highrot}. 
Also, the normal or 
crystal lattice component may have a different velocity than the superfluid component, with the velocity difference proportional to $\left(\nabla\phi- \frac{m}{\hbar}\frac{D\mathbf{u^{ss}}}{Dt}\right)$  where 
$\frac{D\mathbf{u^{ss}}}{Dt}=\frac{\partial\mathbf{u^{ss}}}{\partial t}+\frac{\hbar}{m}\nabla\phi\cdot\nabla\mathbf{u^{ss}}$, giving rise to a kinetic energy term corresponding to mass density of supersolid lattice in $\mathcal{E}_{ph}(\phi)$.  Within the two fluid description, $\rho^{ss}$ is the  density of the superfluid part of the supersolid and $(\rho\delta_{ik}-\rho_{ik}^{ss})$ is the density of the normal (remaining lattice) part of the supersolid lattice, with $\rho$ as the total density of the supersolid.  Also as stated earlier, 
we ignore the associated vortex  lattice effective mass 
in the present set of calculations.

To include the elastic properties of the supersolid and vortex lattice we use free energy of the deformed crystal \cite{landau} such that the strain energy 
$\epsilon_{ik}=\frac{1}{2}\left(\frac{\partial u_{i}}{\partial x_{k}}+\frac{\partial u_{k}}{\partial x_{i}}\right)$. $\lambda_{iklm}$ is a tensor of rank four which relates the strains to the stresses and called as the \textit{elastic modulus tensor}. 

\section{Hydrodynamic Equations for ultracold rotating supersolid}\label{HydEq}
Extremization of the above Lagrangian gives the 
hydrodynamic equations for a rotating supersolid with an embedded vortex lattice and provides the theoretical framework of this paper. These equations are 
\begin{widetext}
\bea \frac{\partial\rho}{\partial t}+\nabla\cdot\left(\rho\frac{\hbar}{m}\nabla\phi\right)+\frac{\partial}{\partial x_{k}}\left[(\rho-\rho^{ss})(\delta_{ik}
-\partial_{k}u_{i}^{ss})\left(\dot{u}_{i}^{ss}-\frac{\hbar}{m}\partial_{i}\phi\right)\right]&=& 0\label{hyd1}
\\
m\left(\frac{\partial \mathbf{v_{s}}}{\partial t}+\tilde{\mathbf{\omega}}\times\mathbf{v}_{L}\right) & = & -\frac{\nabla P'}{\rho}\label{hyd2} \\
2\Omega \rho [\hat{z}\times(\mathbf{v}_{L}-\mathbf{v}_{s})] -  \frac{(\lambda^{v}+\mu_{s}^{v})\nabla(\nabla\cdot\mathbf{u}^{v})+\mu_{s}^{v}\nabla^{2}
\mathbf{u}^{v}}{m} & = & 0 \label{hyd3} \\
 m\frac{\partial}{\partial t}\left[(\rho-\rho^{ss})\left(\dot{u}_{i}^{ss}-\frac{\hbar}{m}\partial_{i}\phi\right)\right]
 +  \hbar\frac{\partial}{\partial x_{k}}\left[(\rho-\rho^{ss})\left(\dot{u}_{i}^{ss} -\frac{\hbar}{m}\partial_{i}\phi\right)\partial_{k}\phi\right] &  & \nonumber\\
\mbox{} -\left[(\lambda^{ss}+\mu_{s}^{ss})\partial_{ik}u_{k}^{ss}+\mu_{s}^{ss}\nabla^{2}u_{i}^{ss}\right] & =& 0
\label{hyd4}\eea
\end{widetext}
Eq. (\ref{hyd1}) and Eq. (\ref{hyd2}) correspond to the equations of motion for density 
and phase and implies conservation of mass and momentum respectively. In deriving them, higher order terms containing product of derivatives of different quantities  like  $\frac{\partial}{\partial\rho}(\rho-\rho^{ss})\left(\nabla\phi-\frac{m}{\hbar}\frac{D\mathbf{u}^{ss}}{Dt}\right)^{2}$, $ \frac{(\rho^{2}-1)\delta\rho}{\rho} (\nabla\phi)^{2}$ and $(\nabla\cdot \mathbf{u}^{ss})(\frac{m}{\hbar})^{2}(\frac{D\mathbf{u}^{ss}}{Dt})^{2}$ are neglected.  We perform an averaging over vortex lattice cell to get  $\mathbf{\tilde{\omega}}=2\mathbf{\Omega}+\nabla\times\mathbf{v}_{s}$ as the averaged vorticity, with the time derivative $\dot{\mathbf{u}}^{v}$ giving the velocity of the vortex lattice $\mathbf{v}_{L}$  and  $\mathbf{v_{s}}$ as the averaged superfluid velocity ($\mathbf{v_{s}}=\frac{\hbar}{m}\nabla\phi$). The pressure $P'=\rho\left(T+\int U(|\mathbf{r}-\mathbf{r}'|)\rho(\mathbf{r}')d\mathbf{r}'\right)$, with $T$ as the quantum pressure term, given by $T=\frac{\hbar^{2}}{2m\sqrt{\rho}}\nabla^{2}\sqrt{\rho}$. $T$ shows the quantum mechanical nature as it contains $\hbar^{2}$ explicitly, and hence termed as quantum pressure term.
As pointed out earlier, we assume throughout an isotropic  supersolid lattice, such that the superfluid density tensor $\rho_{ik}^{ss}=\rho^{ss}\delta_{ik}$.  

Eq.(\ref{hyd3}) is obtained by putting 
the force $\mathbf{f} ( \mathbf{f}_{el}^{v}=-\rho\mathbf{\tilde{\omega}}\times(\mathbf{v}_{L}-\mathbf{v}_{s}))$ acting per unit volume of the fluid moving with
velocity $\mathbf{v}_{s}$,  equal to 
the variation of the elastic energy due to vortex displacements given by 
$ f_{el,i}^{v}=-\frac{\delta\mathcal{E}_{el}^{v}}{\delta u_{i}^{v}}=-\frac{\partial}{\partial x_{k}}\left(\frac{\delta\mathcal{E}_{el}^{v}}
{\delta(\partial u_{k}^{v}/\partial x_{i})}\right)$. Eq.(\ref{hyd4}) gives elastic response of the isotropic supersolid crystal lattice. Here $\lambda^{ss,v}=K^{ss,v}-\frac{2}{3}\mu_{s}^{ss,v }$  are respectively 
second Lame coefficient of the supersolid and vortex lattice, and $K^{ss,v}$ and $\mu_{s}^{ss,v}$
are the respective compressibility and shear modulus of these lattices \cite{landau}.

The non-deformed steady state of a supersolid with embedded vortex lattice  is characterized by $\mathbf{u}^{ss}=0, \mathbf{u}^{v}=0$ ( and therefore $\mathbf{u}=0$), and an average density $\bar{\rho}$.  Eqs.(\ref{hyd1}-\ref{hyd4}) are now linearized around such a steady state  in terms of small perturbations  $\delta\rho$, $\delta\phi$, $\nabla\cdot\mathbf{u}^{ss}=\delta J^{ss}$ and $\mathbf{u}^{v}$.  Here, $\delta J^{ss}$ is the elastic compressibilty of the supersolid lattice, and  equation (\ref{rhoexpand}) shows the relation between  $\delta \rho$
and  $\delta J^{ss}$. 
The resulting equations describe the low energy collective excitations of a rotating supersolid and are given by
\begin{widetext}
\bea \frac{\partial\delta\rho}{\partial t}+\rho^{ss}\frac{\hbar}{m}\nabla^{2}\delta\phi+ (\bar{\rho} - \rho^{ss})\frac{\partial}{\partial t}\delta J_{ss}&=& 0\label{dyna1} \\
 \rho^{ss}\left(\frac{\partial \mathbf{v_{s}}}{\partial t}+2\mathbf{\Omega}\times\mathbf{v_{L}}\right) &= &-c_{sm}^{2}\mathbf{\nabla}\delta\rho \label{dyna2}\\
 \bar{\rho} 2\Omega[\hat{z}\times(\mathbf{v_{L}}-\mathbf{v_s})] &= & \frac{(\lambda^{v}+\mu_{s}^{v})\nabla(\nabla\cdot\mathbf{u}^{v})+\mu_{s}^{v}\nabla^{2}
\mathbf{u}^{v} }{m} \label{dyna3m} \\
\frac{\partial}{\partial t}\left(\frac{\partial}{\partial t}\delta J^{ss}-\frac{\hbar}{m}\mathbf{\nabla^{2}}\delta\phi\right)
&= & \frac{(\lambda^{ss}+2\mu_{s}^{ss})}{ m(\bar{\rho}-\rho^{ss})}
\mathbf{\nabla^{2}}\delta J^{ss} \label{dyna4}\eea
\end{widetext}
In eq.(\ref{dyna2}),  $c_{sm}$ is the modified sound velocity, namely 
 $c_{sm}^{2}  =  c_{s}^{2} (\rho^{ss}/\bar{\rho})$, where $c_{s}$ is the usual sound velocity 
that connects the pressure fluctuation $\delta P'$ to density through
$\delta P'  =  mc_{s}^{2}\delta\rho$. $\mathbf{v_{L}}$ and $\mathbf{v_{s}}$ in eqs. (\ref{dyna2}) and (\ref{dyna3m}) are the vortex lattice velocity, and the averaged superfluid velocity respectively. Eq.(\ref{dyna4}) has been obtained after taking 
divergence of the Eq.(\ref{hyd4}) and then performing the linearization, with  
\beq \delta J^{ss} = \nabla\cdot\mathbf{u}^{ss}\eeq
 as the elastic compressibility of the supersolid lattice. 

Apart from the above set of equations, there is another equation that describes  
the decoupled shear waves for the rotating
supersolid system. It is obtained by taking curl of equation (\ref{hyd4}) after expanding in terms of small fluctuations, 
which gives
\beq m(\bar{\rho}-\rho^{ss})\frac{\partial^{2}}{\partial t^{2}}\varpi-\mu_{s}^{ss}\nabla^{2}\varpi=0\label{sheareq}\eeq
where
\beq \varpi=\mathbf{\nabla}\times\mathbf{u}^{ss}\label{curlu}\eeq
This equation (\ref{sheareq}) gives the shear mode velocity which depends on the supersolid density, namely
\beq v_{shear}^{ss}=\sqrt{\frac{\mu_{s}^{ss}}{m(\bar{\rho}-\rho^{ss})}}\label{shearss}\eeq
 It is to note that the shear mode for the supersolid is obtained by taking curl of the equation for elastic response of the supersolid lattice, and the divergence of the same equation is used to calculate the longitudinal modes of the supersolid lattice.


\subsection{ Low energy long wavelength modes from hydrodynamic equations}
In the rapid rotation limit, after setting the reduced trapping potential to zero,
we expand the small fluctuations as 
\bea \delta\rho  & = & \delta\rho(\mathbf{q})~\text{exp}(i\mathbf{q}\cdot \mathbf{r}-i\omega t) \nonumber \\
\mathbf{v_{s}}  & = &  \mathbf{v_{s}}(\mathbf{q})~\text{exp}(i\mathbf{q}\cdot\mathbf{r}-i\omega t) \nonumber \\  
\delta J^{ss}  & =  & \delta J^{ss}(\mathbf{q})~\text{exp}(i\mathbf{q}\cdot\mathbf{r}-i\omega t). \label{fluceq} \eea 
Here, $\mathbf{r}$  and $\mathbf{q}$ are two dimensional position vector in plane normal to rotation
axis and two dimensional wave vector. We decompose superfluid velocity $\mathbf{v_{s}}$ and the vortex lattice velocity $\mathbf{v_{L}}$ in longitudinal and transverse component
in the $\mathbf{q}$ plane. Subsequent algebra in Fourier space express the longitudinal and transverse components
of the superfluid velocity $\{ v_{sq},v_{st} \}$ in terms of the longitudinal and transverse component of the vortex 
lattice velocity $\{v_{Lq},v_{Lt}\}$.

In terms of longitudinal and transverse components of various velocity, we get the following equations for determining the dispersion relation :
\beq i\omega v_{Lt}-v_{Lq}\left(\frac{c_{vl}^{2}q^{2}}{2\Omega}+2\Omega\right)=0 \label{disp1}\eeq
\beq i\omega v_{Lq}+v_{Lt}\left[\frac{c_{vs}^{2}q^{2}}{2\Omega}+2\Omega\frac{\omega^{2}}{(\omega^{2}-c_{sm}^{2}q^{2})}\right]-
i\frac{c_{sm}^{2}q\omega^{2}(1-\bar{\rho}/\rho^{ss})}{(\omega^{2}-c_{sm}^{2}q^{2})}\delta J^{ss}=0 \label{disp2}\eeq
\beq \left[m\omega^{2}(\bar{\rho}-\rho^{ss})\frac{\omega^{2}-c_{s}^{2}q^{2}}{\omega^{2}-c_{sm}^{2}q^{2}}-q^{2}(\lambda^{ss}+2\mu_{s}^{ss})\right]\delta J^{ss}+im(\bar{\rho}-\rho^{ss})
2\Omega \frac{\omega^{2}}{\omega^{2}-c_{sm}^{2}q^{2}}q v_{Lt}=0\label{disp3}\eeq
where we have labelled the longitudinal and shear parts of vortex lattice velocities by
\beq c_{vl}^{2} =\frac{\lambda^{v}+2\mu_{s}^{v}}{m\bar{\rho}}; c_{vs}^{2}=\frac{\mu_{s}^{v}}{m\bar{\rho}}\label{longshearvl}\eeq
%
%

Substituting these relations in the Fourier transformed form of Eq.(\ref{dyna3m}) and Eq.(\ref{dyna4}), we can finally write the linearized equations in matrix form as 
\begin{widetext}
\beq \begin{bmatrix} 
i \omega  &  -\left(\frac{c_{vl}^{2}q^{2}}{2\Omega}+2\Omega\right) & 0 \\
\left[\frac{c_{vs}^{2}q^{2}}{2\Omega}+2\Omega\frac{\omega^{2}}{(\omega^{2}-c_{sm}^{2}q^{2})}\right] & 
i \omega & - i\frac{(c_{sm}^{2}-c_{s}^{2})q\omega^{2}}{(\omega^{2}-c_{sm}^{2}q^{2})} \\
 \frac{2im \Omega\omega^{2} (\bar{\rho}-\rho^{ss}) }{\omega^{2}-c_{sm}^{2}q^{2}}q & 0 & 
 \left[m\omega^{2}(\bar{\rho}-\rho^{ss})\frac{\omega^{2}-c_{s}^{2}q^{2}}{\omega^{2}-c_{sm}^{2}q^{2}}-q^{2}(\lambda^{ss}+2\mu_{s}^{ss})\right] \end{bmatrix}
\begin{bmatrix} v_{Lt} \\ v_{Lq} \\ \delta J^{ss} \end{bmatrix} = 0  \eeq 

It gives the following dispersion equation 
\bea \omega^{6}-\omega^{4}\left(c_{km}^{2}q^{2}+c_{s}^{2}q^{2}+\left(\frac{c_{vl}^{2}q^{2}}{2\Omega}+2\Omega\right)\left( \frac{c_{vs}^{2}q^{2}}{2\Omega}+2\Omega\right)\right)\nonumber\\
\mbox{} +\omega^{2}\left[\left(c_{sm}^{2}q^{2}+\left(\frac{c_{vl}^{2}q^{2}}{2\Omega}+2\Omega\right)\left(\frac{c_{vs}^{2}q^{2}}{2\Omega}+2\Omega\right)\right)c_{km}^{2}q^{2}+\left(\frac{c_{vl}^{2}q^{2}}{2\Omega}+2\Omega\right)\left(\frac{c_{vs}^{2}q^{2}}{2\Omega}\right)(c_{s}^{2}q^{2})\right]\nonumber\\
\mbox{} -\left(\frac{c_{vl}^{2}q^{2}}{2\Omega}q^{2}+2\Omega\right)\left(\frac{c_{vs}^{2}q^{2}}{2\Omega}\right)(c_{sm}^{2}q^{2})(c_{km}^{2}q^{2})&=& 0\nonumber\\
\label{dispersion1}\eea

which in the long wavelength limit finally leads to the dispersion  
\bea \omega^{6}-\omega^{4}(4\Omega^{2}+c_{km}^{2}q^{2}+(c_{vl}^{2}+c_{vs}^{2})q^{2} + c_{s}^{2}q^{2})+ \nonumber \\
\omega^{2}\left( (4\Omega^{2}+c_{sm}^{2}q^{2}+(c_{vl}^{2}+c_{vs}^{2})q^{2})
c_{km}^{2}q^{2} + c_{vs}^{2}q^{2}c_{s}^{2} q^{2} \right )
\mbox{}-(c_{vs}^{2}q^{2})(c_{sm}^{2}q^{2})(c_{km}^{2}q^{2})&=&0\nonumber\\
\label{dispfull1}\eea
\end{widetext}
where the velocities of  longitudinal and  shear modes  are given by 
$ c_{vl}^{2} =\frac{\lambda^{v}+2\mu_{s}^{v}}{m\bar{\rho}}$ and 
$c_{vs}^{2}=\frac{\mu_{s}^{v}}{m\bar{\rho}}$, and $c_{km}=\sqrt{\frac{\lambda^{ss}+2\mu^{ss}}{m(\bar{\rho}-\rho^{ss})}}$.
Eq.(\ref{dispfull1}) describes the dispersion of a rotating supersolid and is one of the main results in this work.

\subsection{Limiting behavior of the collective modes : Recovering the non rotating supersolid and rotating superfluid}

We first show that the dispersion relation (\ref{dispfull1}) reproduces correct limiting behavior.
Dispersion relation \cite{Pomeau, Rica2,newrica} of a non rotating supersolid can be obtained from (\ref{dispfull1}) by setting the condition that for  $\Omega=0$, there is no vortex lattice.
This gives  $\omega^{2} = \frac{1}{2}(c_{km}^{2} + c_{s}^{2})q^{2}\left[1\pm\sqrt{1-\frac{4 c_{sm}^{2}c_{km}^{2}}{(c_{km}^{2}+c_{s}^{2})^{2}}}\right]$ along with a decoupled shear mode through (\ref{sheareq}),
 reproduces the known result for modes for a non-rotating supersolid as calculated in \cite{Pomeau, Rica2,newrica}. It is to point out that for low $q$ values, our analytical results also qualitatively agree with the appearance of two distinct longitudinal modes for a supersolid, in a recent work by Saccani et.al \cite{boninsegni} derived from a microscopic model using quantum Monte carlo method. However, the analytical approach also gives us the third mode which is absent in the quantum Monte carlo calculations \cite{boninsegni}.
We plot these modes in  Fig. \ref{nrss+rss}(a). 


Results for 
rotating superfluid with a vortex lattice can also be obtained from (\ref{dispfull1})
where $\bar{\rho} \rightarrow  \rho^{ss}$ and $c_{sm} \rightarrow c_{s}$ in the absence of any normal component.
Under these circumstances the following things happen. Firstly, the modified elastic wave 
speed due to presence of the normal component drops out of the description. Secondly the modified second sound velocity 
\beq c_{sm}^{2}=c_{s}^{2}\frac{\rho^{ss}}{\bar{\rho}}\label{modssv}\eeq
becomes the second sound velocity $c_{sm}=c_{s}$.

To see how this  limit correctly reproduces the result for a rotating superfluid, 
we separate out in the equation the terms that depend on $c_{km}^{2}$ by writing it as 
\bea [\omega^{6}-\omega^{4}(4\Omega^{2}+ (c_{vl}^{2}+c_{vs}^{2})q^{2} + c_s^{2}q^{2}) + \omega^{2}c_{vs}^{2}q^{2}c_{s}^{2}q^{2}]\nonumber\\
=c_{km}^{2} q^{2} [\omega^{4} -
\omega^{2}(4\Omega^{2}+c_{sm}^{2}q^{2}+(c_{vl}^{2}+c_{vs}^{2})q^{2}) + c_{vs}^{2} q^{2} c_{sm}^{2} q^{2} ] \label{eqref}\eea

Since $c_{km}=\sqrt{\frac{\lambda^{ss}+\mu^{ss}}{m(\bar{\rho}-\rho^{ss})}}$, we now multiply both side of the equation by $\bar{\rho} - \rho^{ss}$ and take the limit $\rho^{ss}  \rightarrow \bar{\rho} $, which makes the left hand side of eq. (\ref{eqref}) to be zero. We also set $c_{sm}=c_{s}$. This yields 
 \beq \omega^{4} -
\omega^{2}(4\Omega^{2}+c_{s}^{2}q^{2}+(c_{vl}^{2}+c_{vs}^{2})q^{2}) + c_{vs}^{2} q^{2} c_{s}^{2} q^{2}
=0 \eeq 

If we take the equilibrium state as a hexagonal isotropic lattice of vortices following standard literature \cite{Baym3, Baym, Sonin1, Sonin}
and assume that the shear mode velocity is much smaller than other mode velocities, 
the corresponding mode frequencies  are given as  (Fig. \ref{nrss+rss}(b)) 
$ \omega_{I}^{2}=\left[c_{s}^{2}q^{2}+4\Omega^{2}+\frac{4(C_{1}+C_{2})}{m\bar{\rho}}\right]$ and 
$\omega_{T}^{2}=\frac{2C_{2}}{m\bar{\rho}}\frac{c_{s}^{2}q^{4}}{\left[c_{s}^{2}q^{2}+4\Omega^{2}+\frac{4(C_{1}+C_{2})}{m\bar{\rho}}\right]}$. 
This agrees with the earlier results \cite{Baym, Sonin}  where $C_{1}$ and $C_{2}$ are given as 
$ \frac{c_{vl}^{2}}{2 \Omega}=\frac{1}{2\Omega}\frac{4C_{1}+2C_{2}}{m\bar{\rho}}; \frac{c_{vs}^{2}}{2 \Omega}=\frac{1}{2\Omega}\frac{2C_{2}}{m\bar{\rho}}$. 

 \subsection{Results and discussion}
We shall now analytically determine and analyse the roots of the dispersion equation (\ref{dispfull1}) for a rotating supersolid.
Even though the general nature of solutions of such cubic (in terms of $\omega^{2}$) equations (\ref{dispfull1}) are quite involved, the above dispersion relation gets simplified when the velocity associated with the shear mode of the vortex lattice is smaller 
compared to the other mode velocities. This criterion is generally met for the rotating ultra cold atomic superfluid \cite{Baym, Baym1} and therefore it is reasonable to assume  a similar condition for the ultra cold atomic supersolid as well. 

In the current case such a condition reads as  $ (c_{vs}^{2}q^{2})(c_{sm}^{2}q^{2})<<(4\Omega^{2}+c_{sm}^{2}q^{2}+(c_{vl}^{2}+c_{vs}^{2})q^{2} + 
\frac{c_{s}^{2}}{c_{km}^{2}} c_{vs}^{2} q^{2})^{2}$. This means that the last term in
(\ref{dispfull1}) can be neglected to get a quadratic  equation of the form $ x^{2} - B'x +C'=0$. Here $x=\omega^{2}$,
$B'= (4\Omega^{2}+c_{km}^{2}q^{2}+(c_{vl}^{2}+c_{vs}^{2})q^{2} +c_{s}^{2}q^{2})$ and $C'=(4\Omega^{2}+c_{sm}^{2}q^{2} 
+(c_{vl}^{2}+c_{vs}^{2})q^{2})c_{km}^2 q^{2} +(c_{vs}^{2}q^{2})(c_{s}^{2}q^{2})$. In the limit of high rotation frequency and low $q$, $ C' < B'^{2}$ and 
the roots can be approximated as $B'$ and $\frac{C'}{B'}$.  Consequently we get two mode frequencies, namely
\bea \omega_{1}^{2} & \simeq & 4\Omega^{2}+c_{km}^{2}q^{2}+c_{s}^{2}q^{2}+ (c_{vl}^{2}+c_{vs}^{2})q^{2}\label{root1} \\
\omega_{2}^{2} & \simeq & \frac{(4\Omega^{2}+c_{sm}^{2}q^{2}+(c_{vl}^{2}+c_{vs}^{2})q^{2})c_{km}^{2}q^{2} +c_{vs}^{2}q^{2}c_{s}^{2}q^{2}}{(4\Omega^{2}+c_{km}^{2}q^{2}
+c_{s}^{2}q^{2}+ (c_{vl}^{2}+c_{vs}^{2})q^{2})}
\label{root2}\eea
There is a decoupled shear mode which also exists alongwith the above two modes, which is given by
\bea  \omega_{3}^{2}&=&\frac{\mu_{s}^{ss}}{m(\bar{\rho}-\rho^{ss})}q^{2}\label{shearmode} \eea
The above three modes provides us the bulk excitation spectrum of the rotating supersolid within this hydrodynamic approximation and forms one of the main findings of the work. 
The first mode (\ref{root1}) is the inertial mode of the
rotating supersolid, which for $\Omega<<c_{km}q$, behaves as a sound wave, while for $\Omega>>c_{km}q$,  the frequency of the mode begins essentially
at $2\Omega$ i.e. it a gapped mode for rotating supersolid for $\Omega>>c_{km}q$. Corresponding inertial modes for rotating superfluids have been calculated and observed experimentally \cite{Coddington}, with $c_s$ as the sound speed.  
In the second mode (\ref{root2}), as can be seen from equation (\ref{root2}), all the three velocities, the supersolid lattice velocity, the superfluid velocity and
the vortex lattice velocity gets coupled. This mode is unique for the case of rotating supersolid case, and can be used to identify and study the co-existing properties of supersolid lattice and the vortex lattice. The third mode (\ref{shearmode}) is decoupled from the other two modes, and it arises due to the existence of supersolid lattice structure, and gives a signature of supersolidity in the system. This mode also appears for the supersolid phase without any rotation, and stays unaffected by the vortex lattice structure co-existing with the supersolid lattice in the present case. 

The existence of the decoupled shear mode in non-rotating (Fig. \ref{nrss+rss} (a)) as well as rotating supersolid (Fig. \ref{nrss+rss} (c)) characterizes the signature of periodic crystalline order embedded with in the superfluid in the super solid phase (be it non-rotating or rotating). However, one can notice the change in the mode frequencies and their dispersion relations from the Fig. \ref{nrss+rss} (a) and (c).  This figure shows the change in the mode frequencies between the rotating super solid with the counterpart non-rotating super solid and rotating superfluid. The appearance of complex modes due to the interplay of vortex lattice and super solid lattice is evident from the equations (\ref{root1}), (\ref{root2}) and (\ref{shearmode}).  All three modes have been plotted in the Fig. \ref{nrss+rss} (c). A more detailed analysis of these modes as symmetric and anti-symmetric combinations of individual modes of vortex lattice and supersolid lattice is given in the next section.

\begin{figure}[htb]
\centering

{%
    \includegraphics[width=0.9\textwidth]{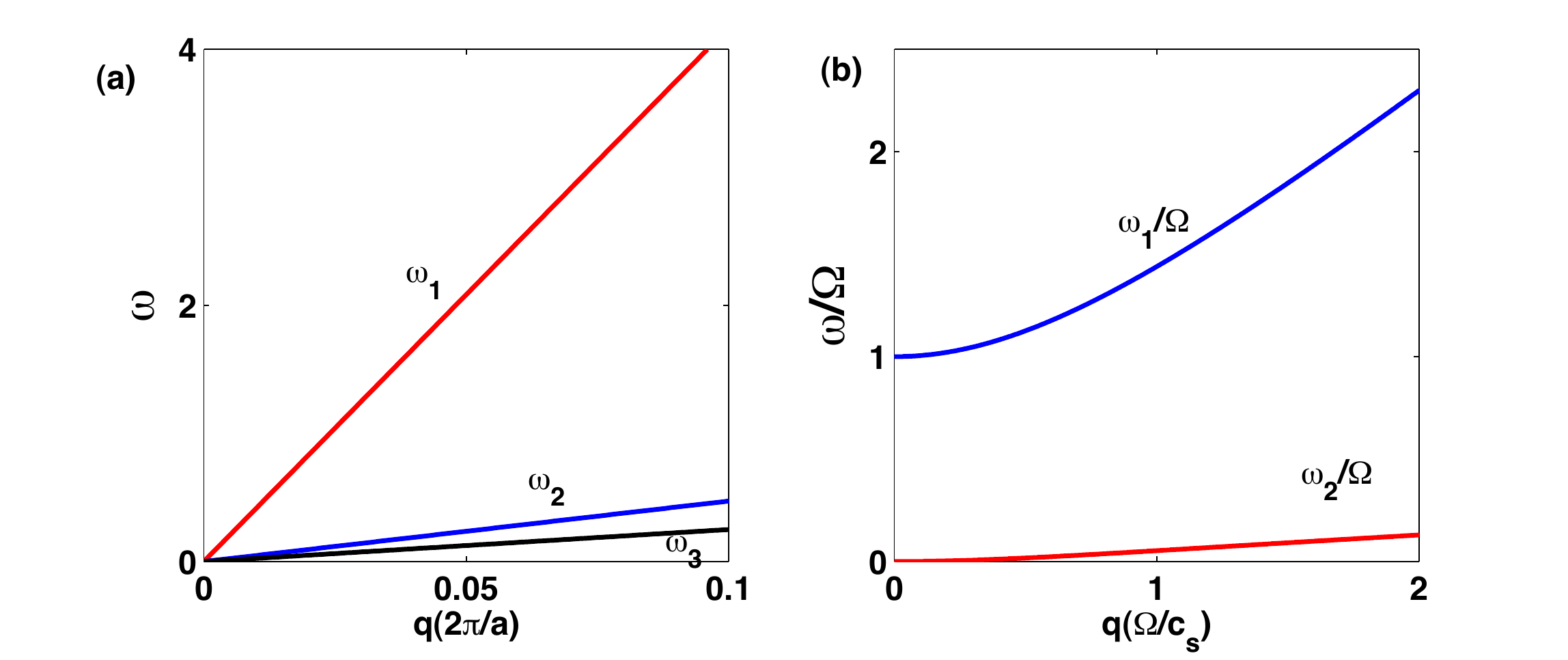}}\\
 
{%
    \includegraphics[width=0.7\textwidth]{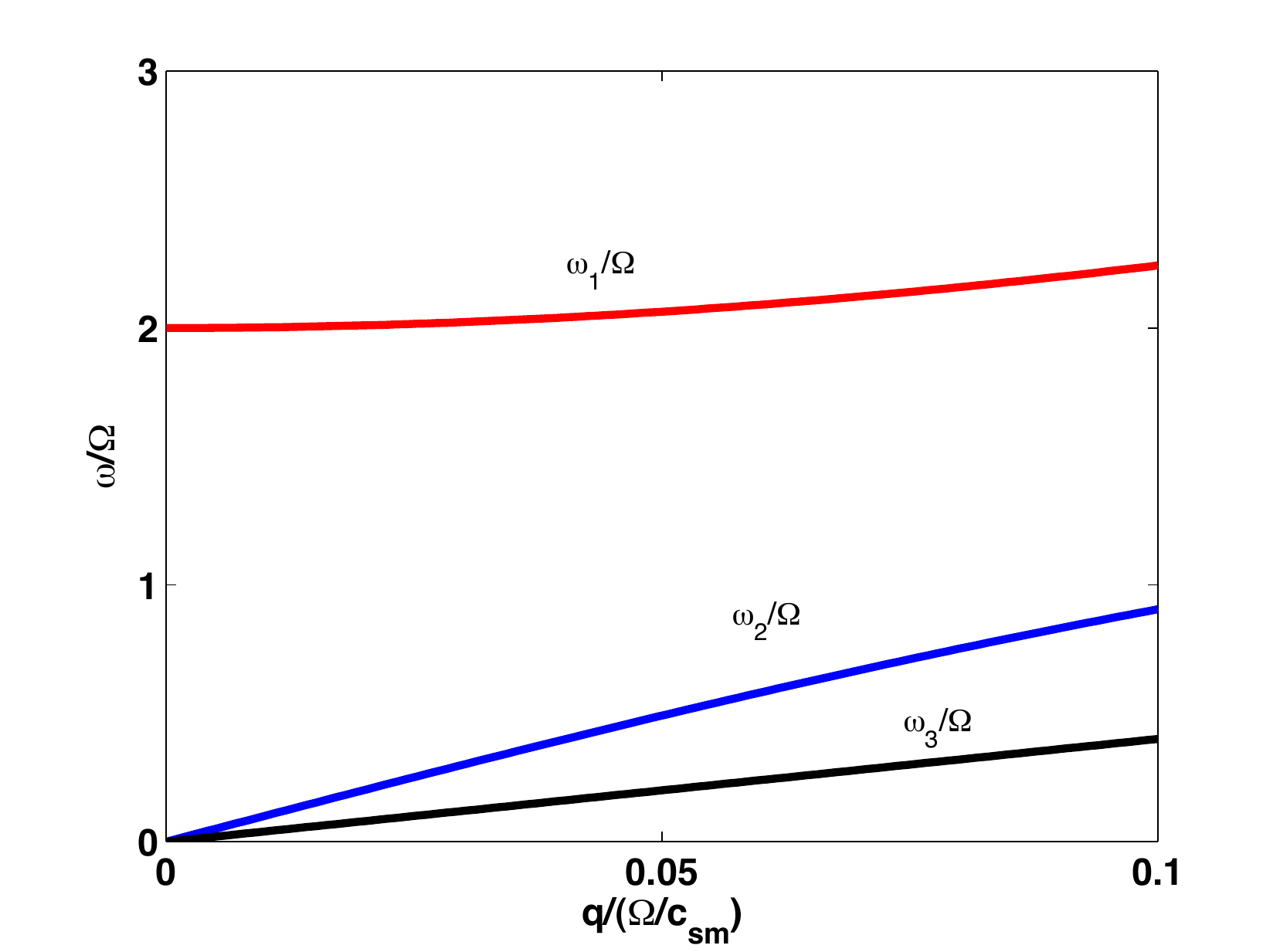}}
    \caption{Dispersion roots for (a) non-rotating supersolid, (b) rotating superfluid and (c) rotating supersolid,  $\omega$ as a function of wave vector $q$, scaled according to the respective cases. It is to note that the parameters
for the elastic wave velocity $c_{k}$, superfluid velocity $c_{s}$ are taken from ref.
\cite{boninsegni} with  $f_{ss}=0.3$.  Here, $ f_{ss}$ is the amount of superfluid fraction in the supersolid, and the value of vortex lattice velocity $c_{vl}$ and $c_{vs}$ have been taken from \cite{Baym}.
  }
 \label{nrss+rss}
\end{figure}

%

\subsection{Symmetric and antisymmetric combination of modes of an rotating ultracold supersolid:}
To understand the significance of the mode frequencies, we can rewrite the first collective mode 
frequency (\ref{root1})  $\omega_{1}^{2} = \omega_{vl}^{2} +  \omega_{ss}^2$ with $\omega_{vl}^{2}  =   [4\Omega^{2}+(c_{vl}^{2}+c_{vs}^{2})q^{2}] $, $\omega_{ss}^{2}  =  (c_{km}^{2}+c_{s}^{2})q^{2}$. This is a  symmetric combination of the square of modes  corresponding 
to the vortex lattice and the supersolid lattice. In the limit of fast rotation and small wave vector,  
the  second mode (\ref{root2}) can be written as
\bea \omega_{2}^{2}   & = &  c_{km}^{2} q^{2} 
 \frac{4\Omega^{2}+c_{sm}^{2}q^{2}+(c_{vl}^{2}+c_{vs}^{2})q^{2} +c_{vs}^{2}q^{2}\frac{c_{s}^{2}}{c_{km}^{2}}}{(4\Omega^{2}+c_{km}^{2}q^{2}
+c_{s}^{2}q^{2}+ (c_{vl}^{2}+c_{vs}^{2})q^{2})} \nonumber \\
& \approx &  \frac{c_{km}^{2} q^{2}}{\omega_{vl}^{2}} \left[ \omega_{vl}^{2} - \omega_{ss}^{2}  + \left( c_{sm}^{2} + \frac{c_{s}^{2}}{c_{km}^{2}} c_{vs}^{2}\right)q^{2} \right ]
\eea 
 To understand this mode, in the limit of fast rotation and small $q$ behavior,  we set the simplifying assumption 
$   (c_{sm}^{2} + \frac{c_{s}^{2}}{c_{km}^{2}} c_{vs}^{2})q^{2} \ll 4 \Omega^{2}$ and hence the left hand side term can be dropped. This sets $\omega_{2}^{2} = 
\frac{c_{km}^{2}q^{2}}{\omega_{vl}^{2}}(\omega_{vl}^{2} - \omega_{ss}^{2})$. The same limit also ensures that 
the preceeding expression is always positive and will not lead to any instability. This expression 
gives an antisymmetric coupling between the square of modes of vortex and the supersolid lattice. 
For a more realistic situation one can readily calculate the 
modification of this expression by including  neglected terms. The square of normal modes frequencies $\omega_{1}^{2}$ 
and $\omega_{2}^{2}$ can therefore be interpreted as symmetric and antisymmetric  combination of square of modes corresponding to vortex lattice and supersolid lattice.
Since we do not have the expression of the corresponding eigenvectors, it is not possible to comment conclusively on the type of coupling between the oscillation of the vortex lattice and supersolid lattice in the real space corresponding to such modes.  
Nevertheless, the occurrence of such modes indeed signifies a coupled motion of  the supersolid and vortex 
lattice.
This coupled motion took place for the lowest order hydrodynamic Lagrangian (\ref{Lageff}) where 
there is no direct coupling between two lattice displacement fields,  $\mathbf{u}^{ss}$ and $\mathbf{u}^{v}$. 

Thus the general nature of our results within hydrodynamic approximations suggests its applicability to minimally provide signatures for the supersolidity in cold atomic systems.  We hope this can be tested with  more detailed numerical investigations 
with specific microscopic models in future. The same experimental 
techniques \cite{Coddington, mizushima} used to study the collective oscillation of vortex lattices in rapidly rotating  superfluid  may  be implemented here. Namely one  need to perturb  the system to 
induce a deformation in the co-existing supersolid and vortex lattice for the case of rotating supersolid. The oscillations of these lattices under these perturbations can be observed using the TOF expansion technique and the information about the modes can be extracted and compared with the well known Tkachenko modes for rotating superfluid. This will possibly  provide a route for the confirmation of supersolidity in such ultra cold atomic condensates. 

RS gratefully acknowledges the support provided by CSIR, New Delhi, India.
\appendix
\label{app1}
\section{Derivation of effective Lagrangian for rotating supersolid}\label{app11}
\subsection{Homogenization technique for long wave effective Lagrangian}
Here we derive the coupled equations for the three fields $\rho(\mathbf{r},t)$, $\mathbf{u}(\mathbf{r},t) $ which is a function of $\mathbf{u}^{ss}(\mathbf{r},t)$ and $\mathbf{u}^{v}
(\mathbf{r},t)$ and $\phi(\mathbf{r},t)$, following the method called Homogenization. This technique splits the long wave
behavior of various parameters and the short range periodic dependence on the lattice parameters \cite{Rica2}.

We use the ansatz for density and phase as
\beq n(\mathbf{r},t)=\rho_{0}(\mathbf{r}-\mathbf{u}(\mathbf{r},t)|\rho(\mathbf{r},t))+\tilde{\rho}
(\mathbf{r}-\mathbf{u}(\mathbf{r},t),\rho,t)+... \label{rhoans}\eeq
\beq \Phi(\mathbf{r},t)=\phi(\mathbf{r},t)+\tilde{\phi}(\mathbf{r}-\mathbf{u}(\mathbf{r},t),\rho(\mathbf{r},t),t)+... \label{phians}\eeq
Here, the displacement of the vortex lattice and the supersolid lattice enters the modulated density as $\rho_{0}(\mathbf{r}-\mathbf{u}(\mathbf{r},t)(\mathbf{u}^{ss}(\mathbf{r},t),\mathbf{u}^{v}(\mathbf{r},t))|\rho(\mathbf{r},t))$. Also,  $\phi$, $\mathbf{u} (\mathbf{u}^{ss},\mathbf{u}^{v})$ and $\rho$ are slowly varying fields and $\tilde{\phi}$ and $\tilde{\rho}$ are small
and fast varying periodic functions.

Now we calculate the gradients and time derivatives of various expressions which will be further used in the calculations.
\bea (\nabla n)_{i}&=&
 (\delta_{ik}-\partial_{i}u_{k}^{ss}-\partial_{i}u_{k}^{v})\partial_{k}\rho_{0}+\frac{\partial\rho_{0}}{\partial\rho}\frac{\partial\rho}
{\partial x_{i}}
+(\delta_{ik}-\partial_{i}u_{k}^{ss}-\partial_{i}u_{k}^{v})\partial_{k}\tilde{\rho}+\frac{\partial{\tilde{\rho}}}{\partial\rho}\frac{\partial\rho}
{\partial x_{i}}
\eea
Next,
\bea \partial_{t}\Phi
&=& \partial_{t}\phi-\partial_{t}u_{k}^{ss}\partial_{k}\tilde{\phi}-\partial_{t}u_{k}^{v}\partial_{k}\tilde{\phi}+\partial_{t}\tilde{\phi}
+\frac{\partial \tilde{\phi}}{\partial\rho}\partial_{t}\rho\eea
\bea (\nabla\Phi)_{i}
&=& (\nabla\phi)_{i}+(\delta_{ik}-\partial_{i}u_{k}^{ss}-\partial_{i}u_{k}^{v})\partial_{k}\tilde{\phi}+\frac{\partial\tilde{\phi}}{\partial\rho}
(\nabla \rho)_{i}\eea
Now keeping the relevant contributions for the long-wave description and calculating
\bea n\partial_{t}\Phi&=&(\rho_{0}+\tilde{\rho})\left(\partial_{t}\phi-\partial_{t}u_{k}^{ss}\partial_{k}\tilde{\phi}-\partial_{t}u_{k}^{v}\partial_{k}
\tilde{\phi}+\partial_{t}\tilde{\phi}+\frac{\partial \tilde{\phi}}{\partial\rho}\partial_{t}\rho\right)\nonumber\\
&=&\rho_{0}\partial_{t}\phi-\rho_{0}\partial_{t}u_{k}^{ss}\partial_{k}\tilde{\phi}-\rho_{0}\partial_{t}u_{k}^{v}\partial_{k}\tilde{\phi}+
\partial_{t}\phi\tilde{\rho}+h.o.t, \label{rhodelt}\eea
 \textit{h.o.t} stands for higher order terms through out the calculations.
\bea (\nabla n)^{2}
&=& \left((\delta_{ik}-\partial_{i}u_{k}^{ss}-\partial_{i}u_{k}^{v})\partial_{k}\rho_{0}+\frac{\partial\rho_{0}}{\partial\rho}\frac{\partial\rho}{\partial x_{i}}
\right)^{2}+\left((\delta_{ik}-\partial_{i}u_{k}^{ss}-\partial_{i}u_{k}^{v})\partial_{k}\tilde{\rho}+\frac{\partial{\tilde{\rho}}}{\partial\rho}\frac{\partial \rho}
{\partial x_{i}}\right)^{2}\nonumber\\
\mbox{} & & +2\left((\delta_{ik}-\partial_{i}u_{k}^{ss}-\partial_{i}u_{k}^{v})\partial_{k}\rho_{0}+\frac{\partial\rho_{0}}{\partial\rho}\frac{\partial\rho}
{\partial x_{i}}\right)\left((\delta_{ik}-\partial_{i}u_{k}^{ss}-\partial_{i}u_{k}^{v})\partial_{k}\tilde{\rho}+\frac{\partial{\tilde{\rho}}}{\partial\rho}
\frac{\partial \rho}{\partial x_{i}}\right)\nonumber\\
\label{rhosq}\eea
We calculate this quantity (\ref{rhosq}) term by term as follows :

\textbf{Term 1}:
\bea \left((\delta_{ik}-\partial_{i}(u_{k}^{ss}+u_{k}^{v}))\partial_{k}\rho_{0}+\frac{\partial\rho_{0}}{\partial\rho}\frac{\partial\rho}
{\partial x_{i}}\right)^{2}
&=& [(\delta_{ik}+(\partial_{i}(u_{k}^{ss}+u_{k}^{v}))^{2}-2\delta_{ik}\partial_{i}(u_{k}^{ss}+u_{k}^{v})]\left(\frac{\partial\rho_{0}}
{\partial x_{k}}\right)^{2}
\nonumber\\
&=&(\delta_{ik}-2\partial_{i}u_{k}^{ss}-2\partial_{i}u_{k}^{v}+\partial_{l}u_{i}^{v}\partial_{l}u_{k}^{v}+\partial_{l}u_{i}^{ss}\partial_{l}u_{k}^{ss})
\frac{\partial\rho_{0}}{\partial x_{i}}\frac{\partial\rho_{0}}{\partial x_{k}}\nonumber\\
&=& (\delta_{ik}+2\epsilon_{ik}^{s}+2\epsilon_{ik}^{v})\frac{\partial\rho_{0}}{\partial x_{i}}\frac{\partial\rho_{0}}{\partial x_{k}}+h.o.t
\label{1stpart}\eea
where
\beq \epsilon_{ik}^{s}=\frac{1}{2}(\partial_{i}u_{k}^{ss}+\partial_{k}u_{i}^{ss})+\frac{1}{2}\partial_{l}u_{i}^{ss}\partial_{l}u_{k}^{ss}\label{strain1}\eeq
is the strain tensor for supersolid lattice and,
\beq \epsilon_{ik}^{v}=\frac{1}{2}(\partial_{i}u_{k}^{v}+\partial_{k}u_{i}^{v})+\frac{1}{2}\partial_{l}u_{i}^{v}\partial_{l}u_{k}^{v}\label{strain2}\eeq
is the strain tensor for vortex lattice \cite{landau}.

\textbf{Term 2}
\bea (\delta_{ik}-(\partial_{i}u_{k}^{ss}+\partial_{i}u_{k}^{v}))^{2}\left(\frac{\partial\tilde{\rho}}{\partial x_{k}}\right)^{2}
+\left(\frac{\partial\tilde{\rho}}
{\partial\rho}\frac{\partial\rho}{\partial x_{i}}\right)+2(\delta_{ik}-(\partial_{i}u_{k}^{ss}+\partial_{i}u_{k}^{v}))
\frac{\partial\tilde{\rho}}{\partial x_{k}}
\frac{\partial\tilde{\rho}}{\partial \rho}\frac{\partial\rho}{\partial x_{i}}\nonumber\eea
\bea &=&\delta_{ik}\left(\frac{\partial\tilde{\rho}}{\partial x_{k}}\right)^{2}+h.o.t \nonumber\\
&=&(\partial_{i}\tilde{\rho})^{2}+h.o.t \label{2ndpart}\eea
As mentioned earlier, in order to keep the relevant terms for long wavelength description, terms which are quadratic in fast varying variable $\tilde{\rho}$ multiplied by other derivatives are ignored.

\textbf{Term 3}
\bea 2\left((\delta_{ik}-\partial_{i}u_{k}^{ss}-\partial_{i}u_{k}^{v})\partial_{k}\rho_{0}+\frac{\partial\rho_{0}}{\partial\rho}\frac{\partial\rho}
{\partial x_{i}}\right)\left((\delta_{ik}-\partial_{i}u_{k}^{ss}-\partial_{i}u_{k}^{v})\partial_{k}\tilde{\rho}+\frac{\partial{\tilde{\rho}}}{\partial\rho}
\frac{\partial \rho}{\partial x_{i}}\right)\nonumber\\
= 2(\delta_{ik}-\partial_{i}u_{k}^{ss}-\partial_{i}u_{k}^{v})^{2}\frac{\partial\rho_{0}}{\partial x_{k}}\frac{\partial\tilde{\rho}}
{\partial x_{k}}+h.o.t
\simeq 2(\delta_{ik}+2\epsilon_{ik}^{s}+2\epsilon_{ik}^{v})\partial_{k}\rho_{0}\partial_{k}\tilde{\rho}\label{3rdpart}\eea

Substituting equations (\ref{1stpart}),(\ref{2ndpart}),(\ref{3rdpart}) into equation (\ref{rhosq}), we get
\bea (\nabla n)^{2}&=&(\delta_{ik}+2\epsilon_{ik}^{s}+2\epsilon_{ik}^{v})\partial_{i}\rho_{0}\partial_{k}\rho_{0}
+2(\delta_{ik}+2\epsilon_{ik}^{s}+2\epsilon_{ik}^{v})\partial_{k}\rho_{0}\partial_{k}\tilde{\rho}+ (\partial_{i}\tilde{\rho})^{2}+h.o.t\nonumber\\
\label{rhosq2}\eea
Next we calculate
\bea (\nabla\Phi)^{2}&=&\left[(\nabla\phi)^{2}+(\delta_{ik}-(\partial_{i}u_{k}^{ss}+\partial_{i}u_{k}^{v}))\frac{\partial\tilde{\phi}}{\partial x_{k}}
+\frac{\partial\tilde{\phi}}{\partial\rho}(\nabla\rho)\right]^{2}\nonumber\\
&=& (\partial_{i}\phi)^{2}+(\nabla\tilde{\phi})^{2}+2(\delta_{ik}-(\partial_{i}u_{k}^{ss}+\partial_{i}u_{k}^{v}))\partial_{i}\phi\partial_{k}\tilde{\phi}
+h.o.t\label{rhodelphi2}\eea
The higher order terms are the terms quadratic in fast varying variable $\tilde{\phi}$ multiplied by other derivatives, which we again neglect in
the long wavelength description. The next term is
\bea n\nabla\Phi&=& (\rho_{0}+\tilde{\rho})\left((\nabla\phi)+(\delta_{ik}-\partial_{i}u_{k}^{ss}-\partial_{i}u_{k}^{v})\partial_{k}\tilde{\phi}
+\frac{\partial\tilde{\phi}}{\partial \rho}(\nabla\rho)\right)\nonumber\\
&=&\rho_{0}\nabla\phi+\rho_{0}(\delta_{ik}-\partial_{i}u_{k}^{ss}-\partial_{i}u_{k}^{v})\frac{\partial\tilde{\phi}}{\partial x_{k}}+\rho_{0}
\frac{\partial\tilde{\phi}}{\partial\rho}(\nabla\rho)+h.o.t \eea
Before going into the calculation for the non-local interaction term, we calculate the $1$st and $2$nd term of Lagrangian (\ref{lagrhophi}) and label
their contribution to the corresponding energy part of the Lagrangian, which is explained later.

In the $1st$ term $\textit{L}_{1}$ we use equation (\ref{rhodelt}) and get the following expression
\bea \textit{L}_{1} & = &-\int\hbar n\frac{\partial\Phi}{\partial t}d\mathbf{r}\nonumber\\
&=& \underbrace{-\int \hbar\partial_{t}\phi\rho_{0}d\mathbf{r}}_{L_{\phi}}+\underbrace{\int \hbar\rho_{0}\partial_{t}u_{k}^{ss}\partial_{k}
\tilde{\phi}d\mathbf{r}}_{L_{\tilde{\phi}}}\nonumber\\
\mbox{} & & +\underbrace{\int\hbar\rho_{0}\partial_{t}u_{k}^{v}
\partial_{k}\tilde{\phi}d\mathbf{r}}_{L_{\tilde{\phi}}}
-\underbrace{\int \hbar\partial_{t}\Phi\tilde{\rho}d\mathbf{r}}_{neglected}\label{lag1}\eea
$2nd$ term of Lagrangian (\ref{lagrhophi}) is calculated using (\ref{rhodelphi2}) as
\bea \textit{L}_{2} &=&-\int\frac{\hbar^{2}}{2m}n(\nabla\Phi)^{2}d\mathbf{r}\nonumber\\
\mbox{} & & +\tilde{\rho}(\partial_{i}\phi)^{2}+\tilde{\rho}(\nabla\tilde{\phi})^{2}+2\tilde{\rho}(\delta_{ik}-\partial_{i}u_{k}^{ss
}-\partial_{i}u_{k}^{v})
\partial_{i}\phi\partial_{k}\tilde{\phi}]
d\mathbf{r}\nonumber\\
&=&-\underbrace{\frac{\hbar^{2}}{2m}\int (\nabla\phi)^{2}\rho_{0}(\mathbf{r})d\mathbf{r}}_{L_{\phi}}
+\underbrace{\frac{\hbar^{2}}{m}\int(\partial_{i}u_{k}^{ss}+\partial_{i}u_{k}^{v})\partial_{i}
\phi\partial_{k}\tilde{\phi}\rho_{0}(\mathbf{r})d\mathbf{r}}_{L_{\tilde{\phi}}}\nonumber\\
\mbox{} & & -\underbrace{\frac{\hbar^{2}}{m}\int\nabla\phi\cdot\nabla\tilde{\phi}\rho_{0}(\mathbf{r})d\mathbf{r}}_{L_{\tilde{\phi}}}
-\underbrace{\frac{\hbar^{2}}{2m}\int(\nabla\tilde{\phi})^{2}\rho_{0}
(\mathbf{r})d\mathbf{r}}_{L_{\tilde{\phi}}}
\label{lag2} \eea

Considering the non-local term now, given by
\beq \textit{N}(\rho(\mathbf{r}),\rho(\mathbf{r'
}))=\frac{1}{2}\int U(|\mathbf{r}-\mathbf{r'}|)\rho(\mathbf{r}-\mathbf{u}^{ss}(\mathbf{r})-\mathbf{u}^{v}(\mathbf{r}))
\rho(\mathbf{r'}-\mathbf{u}^{ss}(\mathbf{r'})-\mathbf{u}^{v}(\mathbf{r'}))d\mathbf{r}d\mathbf{r'}\label{nonlocal}\eeq
\textbf{STEPS}

1) Using the change of variables, $\mathbf{R}=\mathbf{r}-\mathbf{u}(\mathbf{r})$ and, $\mathbf{R'}=\mathbf{r'}-\mathbf{u}(\mathbf{r'})$, we can determine
\beq d\mathbf{R}=d(\mathbf{r}-(\mathbf{u}^{ss}(\mathbf{r})+\mathbf{u}^{v}(\mathbf{r})))\nonumber\eeq
with
\beq (d\mathbf{R})_{i}=d(\mathbf{r}-(\mathbf{u}^{ss}(\mathbf{r})+\mathbf{u}^{v}(\mathbf{r})))_{i}=\left(dx_{i}-\frac{\partial (u^{ss}_{i}+u^{v}_{i})}
{\partial x_{k}}dx_{k}\right)\nonumber\eeq
which implies
\bea |d\mathbf{R}|^{2}&=&(d\mathbf{R})_{i}\cdot(d\mathbf{R})_{i}\nonumber\\
&=& (\delta_{ik}-2(\partial_{k}u_{i}^{ss}+\partial_{k}u_{i}^{v})+(\partial_{k}u_{i}^{ss}+\partial_{k}u_{i}^{v})(\partial_{k}u_{i}^{ss}+\partial_{k}u_{i}^{v}))
dx_{i}dx_{k}\nonumber\\
&=& (\delta_{ik}+2\epsilon_{ik}^{s}+2\epsilon_{ik}^{v})dx_{i}dx_{k}\label{dRexp}\eea
with strain tensors $\epsilon_{ik}^{s}$ and $\epsilon_{ik}^{v}$ defined in equations (\ref{strain1}) and (\ref{strain2}). Similarly,
\beq |d\mathbf{R'}|^{2}=(\delta_{ik}+2\epsilon_{ik'}^{s}+2\epsilon_{ik'}^{v})dx'_{i}dx'_{k}\eeq

2) Any integral with argument $\mathbf{r}-\mathbf{u}(\mathbf{r})$ may be transformed to
\bea \int Q(\mathbf{r}-\mathbf{u}(\mathbf{r}))d\mathbf{r}&=&\int Q(\mathbf{R})d\mathbf{r}\nonumber\\
&=& \int Q(\mathbf{R})\frac{d\mathbf{R}}{\sqrt{det(\delta_{ik}+2\epsilon_{ik}^{s}+2\epsilon_{ik}^{v})}}\label{step2Q}\\
&\simeq& \int Q(\mathbf{r})(1-\epsilon_{kk}^{s}-\epsilon_{kk}^{v})d\mathbf{r}\label{Q}\eea
The step (\ref{step2Q}) in equation (\ref{Q}) is obtained by using equation (\ref{dRexp}).

3) Relative distance
\bea \Delta\mathbf{R}&=&\mathbf{R}-\mathbf{R'}\nonumber\\
&=&\Delta\mathbf{r}-\Delta (\mathbf{u}^{v}(\mathbf{r})+ \mathbf{u}^{ss}(\mathbf{r}))\label{DeltaR}\eea
where $\Delta\mathbf{r}=\mathbf{r}-\mathbf{r'}$. Above equation (\ref{DeltaR}) implies
\bea |\Delta\mathbf{R}|^{2}&\simeq&|\Delta\mathbf{r}|^{2}+|\Delta \mathbf{u}(\mathbf{r})|^{2}-2\Delta\mathbf{r}\cdot\Delta \mathbf{u}(\mathbf{r})\nonumber\\
&=&|\Delta\mathbf{r}|^{2}+2\epsilon_{ik}^{s}\Delta x_{i}\Delta x_{k}+2\epsilon_{ik}^{v}\Delta x_{i}\Delta x_{k}\nonumber\eea
Thus,
\beq |\Delta\mathbf{r}|\simeq \Delta\mathbf{R}-\frac{(\epsilon_{ik}^{s}+\epsilon_{ik}^{v})\Delta x_{i}\Delta x_{k}}{\Delta\mathbf{R}}\label{moddeltar}\eeq

The final result of non-local term given by (\ref{nonlocal}) thus can be calculated as
\bea \textit{N}(\rho(\mathbf{r}),\rho(\mathbf{r'}))& = &\frac{1}{2}\int U\left(|\Delta \mathbf{R}|-(\epsilon_{ik}^{s}+\epsilon_{ik}^{v})\frac{\Delta x_{i}
\Delta x_{k}}{|\Delta\mathbf{R}|}+...\right)
\frac{\rho(\mathbf{R})d\mathbf{R}}{\sqrt{det(\delta_{ik}+2\epsilon_{ik}^{s}+2\epsilon_{ik}^{v})}}\nonumber\\
& & \times\frac{\rho(\mathbf{R'})d\mathbf{R'}}{\sqrt{det(\delta_{ik}+2\epsilon_{ik'}^{s}+2\epsilon_{ik'}^{v})}}\nonumber\\
&=&\frac{1}{2}\int (1-(\epsilon_{ll}^{s}+\epsilon_{ll}^{s'})-(\epsilon_{ll}^{v}+\epsilon_{ll}^{v'}))(U(|\mathbf{r}-\mathbf{r'}|)-
(\epsilon_{ik}^{s}+\epsilon_{ik}^{v})
f_{ik}(\mathbf{r}-\mathbf{r'})+...)\nonumber\\
& & \rho(\mathbf{r})\rho(\mathbf{r'})d\mathbf{r}d\mathbf{r'}
\label{Uterm}\eea
Here,
$f_{ik}(\mathbf{r}-\mathbf{r'})=(x_{i}-x_{i'})(x_{k}-x_{k'})\frac{U(\mathbf{r}-\mathbf{r'})}{|\mathbf{r}-\mathbf{r'}|}$.

The $1$st and $2$nd term of the Lagrangian are already calculated. Here we determine the $3$rd and $4$th terms by substituting
the ansatz in equations (\ref{rhoans}) and (\ref{phians}) in the Lagrangian (\ref{lagrhophi}). $3rd$ term of the Lagrangian (\ref{lagrhophi}) is calculated using
(\ref{rhosq2}) and (\ref{Q}) as
\bea \textit{L}_{3} &=&-\frac{\hbar^{2}}{2m}\int\frac{1}{4n}(\nabla n)^{2}d\mathbf{r}\nonumber\\
&=& -\frac{\hbar^{2}}{2m}\int[(\delta_{ik}+2\epsilon_{ik}^{s}+2\epsilon_{ik}^{v})\partial_{i}\rho_{0}\partial_{k}\rho_{0}+2
(\delta_{ik}+2\epsilon_{ik}^{s}+2\epsilon_{ik}^{v})\partial_{i}\rho_{0}\partial_{k}\tilde{\rho}+(\partial_{i}\tilde{\rho})^{2}+h.o.t]\nonumber\\
\mbox{} & & \cdot\left(\frac{1}{4\rho_{0}(\mathbf{r})}+\frac{\tilde{\rho}(\mathbf{r})}
{4\rho_{0}^{2}(\mathbf{r})}\right)
(1-\epsilon_{kk}^{s}-\epsilon_{kk}^{v})d\mathbf{r}\nonumber\eea
\bea &=& -\underbrace{\frac{\hbar^{2}}{8m}\int\frac{(\nabla\rho_{0})^{2}}{\rho_{0}}d\mathbf{r}}_{L_{\rho}}
+\underbrace{\frac{\hbar^{2}}{8m}\int\frac{(\nabla\rho_{0})^{2}}{\rho_{0}}(\epsilon_{kk}^{s}+\epsilon_{kk}^{v})
d\mathbf{r}}_{L_{u}}\nonumber\\
\mbox{} & & -\underbrace{\frac{\hbar^{2}}{4m}\int (\epsilon_{ik}^{s}+\epsilon_{ik}^{v})\frac{\partial_{i}\rho_{0}\partial_{k}\rho_{0}}{\rho_{0}}
(1-\epsilon_{ll}^{s}-\epsilon_{ll}^{v})d\mathbf{r}}_{L_{u}}-\underbrace{\frac{\hbar^{2}}{4m}\int \frac{\partial_{i}\rho_{0}\partial_{i}\tilde{\rho}}
{\rho_{0}}d\mathbf{r}}_{L_{\tilde{\rho}}}\nonumber\\
\mbox{} & & -\underbrace{\frac{\hbar^{2}}{2m}\int(\epsilon_{ik}^{s}+\epsilon_{ik}^{v})\frac{\partial_{i}\rho_{0}\partial_{i}\tilde{\rho}}
{\rho_{0}}d\mathbf{r}}_{L_{\tilde{\rho}}}
-\underbrace{\frac{\hbar^{2}}{8m}\int\frac{(\nabla\tilde{\rho})^{2}}{\rho_{0}}d\mathbf{r}+\frac{\hbar^{2}}{8m}\int\frac{(\nabla\rho_{0})^{2}}
{\rho_{0}^{2}}\tilde{\rho}d\mathbf{r}}_{L_{\tilde{\rho}}}+...\nonumber\\
\label{lag3}\eea
Now, the $4$th term of the Lagrangian (\ref{lagrhophi}) is calculated using (\ref{Uterm}) as
\bea \textit{L}_{4} &=&-\frac{1}{2}\int U(|\mathbf{r}-\mathbf{r'}|)n(\mathbf{r})n(\mathbf{r'})d\mathbf{r}d\mathbf{r'}\nonumber\\
&=& -\frac{1}{2}\int(1-(\epsilon_{ll}^{s}+\epsilon_{ll}^{s'})-(\epsilon_{ll}^{v}+\epsilon_{ll}^{v'}))(U(|\mathbf{r}-\mathbf{r'}|)-(\epsilon_{ik}^{s}+\epsilon_{ik}^{v})
f_{ik}(\mathbf{r}-\mathbf{r'})+...)\nonumber\\
\mbox{} & & \cdot(\rho_{0}(\mathbf{r})\rho_{0}(\mathbf{r'})+\rho_{0}(\mathbf{r})\tilde{\rho}(\mathbf{r'})+\tilde{\rho}(\mathbf{r})\rho_{0}(\mathbf{r'})+
\tilde{\rho}(\mathbf{r})\tilde{\rho}(\mathbf{r'}))d\mathbf{r}d\mathbf{r'}\nonumber\eea
\bea &=& -\underbrace{\frac{1}{2}\int U(|\mathbf{r}-\mathbf{r'}|)\rho_{0}(\mathbf{r})\rho_{0}(\mathbf{r'})d\mathbf{r}d\mathbf{r'}}
_{L_{\rho}}-\underbrace{\frac{1}{2}\int U(|\mathbf{r}-\mathbf{r'}|)
(\rho_{0}(\mathbf{r})\tilde{\rho}(\mathbf{r'})+\tilde{\rho}(\mathbf{r})\rho_{0}(\mathbf{r'}))d\mathbf{r}d\mathbf{r'}}_{L_{\tilde{\rho}}}\nonumber\\
\mbox{} & & -\underbrace{\frac{1}{2}\int U(|\mathbf{r}-\mathbf{r'}|)\tilde{\rho}(\mathbf{r})\tilde{\rho}(\mathbf{r'})d\mathbf{r}d\mathbf{r'}}
_{L_{\tilde{\rho}}}\nonumber\\
\mbox{} & & -\underbrace{\frac{1}{2}\int[(\epsilon_{ik}^{s}
+\epsilon_{ik}^{v})
f_{ik}(\mathbf{r}-\mathbf{r'})](\rho_{0}(\mathbf{r})\tilde{\rho}(\mathbf{r'})+\tilde{\rho}(\mathbf{r})\rho_{0}(\mathbf{r'}))d\mathbf{r}d\mathbf{r'}
}_{L_{\tilde{\rho}}}\nonumber\\
\mbox{} & & -\underbrace{\frac{1}{2}\int[((\epsilon_{ll}^{s}+\epsilon_{ll}^{s'})+(\epsilon_{ll}^{v}+\epsilon_{ll}^{v'}))
f_{ik}(\mathbf{r}-\mathbf{r'})](\rho_{0}(\mathbf{r})\tilde{\rho}(\mathbf{r'})+\tilde{\rho}(\mathbf{r})\rho_{0}(\mathbf{r'}))d\mathbf{r}d\mathbf{r'}
}_{L_{\tilde{\rho}}}\label{lag5}
\eea
So, adding and collecting all the terms, we get following five kind of terms
\bea \textit{L}=\textit{L}_{\rho}+\textit{L}_{\phi}+\textit{L}_{u}+\textit{L}_{\tilde{\phi}}+\textit{L}_{\tilde{\rho}} \eea
(1) $\textit{L}_{\rho}$ is the \textbf{internal energy part}, which only depends on $\rho_{0}(\mathbf{r})$ which is slowly varying, and is given by
\beq \textit{L}_{\rho}=-\frac{\hbar^{2}}{8m}\int\frac{(\nabla\rho_{0})^{2}}{\rho_{0}}d\mathbf{r}-\frac{1}{2}\int U(|\mathbf{r}-\mathbf{r'}|)
\rho_{0}(\mathbf{r})\rho_{0}(\mathbf{r'})d\mathbf{r}d\mathbf{r'}\eeq

(2) $\textit{L}_{\phi}$ is the \textbf{hydrodynamical part I}, which mixes the slowly varying phase $\phi(\mathbf{r},t)$ and slowly varying density
$\rho_{0}(\mathbf{r})$, and
is given below
\bea \textit{L}_{\phi}&=&-\int \left(\hbar\partial_{t}\phi+\frac{\hbar^{2}}{2m}(\nabla\phi)^{2}\right)\rho_{0}(\mathbf{r})d\mathbf{r}\label{lagphi1}\eea

The term in the integral is the Lagrangian density and we obtain an average energy density
that depends on parameter $\phi$ only, shown below as
\bea \textit{E}(\phi)&=&\frac{1}{V}\int_{V}\left(\hbar\partial_{t}\phi+\frac{\hbar^{2}}{2m}(\nabla\phi)^{2}\right)\rho_{0}(\mathbf{r})d\mathbf{r}\nonumber\\
&\simeq&\left(\hbar\partial_{t}\phi+\frac{\hbar^{2}}{2m}(\nabla\phi)^{2}\right)\rho\eea

Thus, the equation (\ref{lagphi1}) when averaged directly looks like
\bea \textit{L}_{\phi}&=&-\int \left(\hbar\partial_{t}\phi+\frac{\hbar^{2}}{2m}(\nabla\phi)^{2}\right)\rho d\mathbf{r}\eea

(3) $\textit{L}_{u}$ is the \textbf{elastic part I}, given by
\bea \textit{L}_{u} & = &-\frac{\hbar^{2}}{4m}\int(\epsilon_{ik}^{s}+\epsilon_{ik}^{v})\frac{\partial_{i}\rho_{0}\partial_{k}\rho_{0}}{\rho_{0}}(1-\epsilon_{ll}^{s}
-\epsilon_{ll}^{v})d\mathbf{r}+\frac{\hbar^{2}}{8m}
\int\frac{(\nabla\rho_{0})^{2}}{\rho_{0}}(\epsilon_{kk}^{s}+\epsilon_{kk}^{v})d\mathbf{r}\nonumber\\
& & \mbox{} + \frac{1}{2}\int ((\epsilon_{ik}^{s}+\epsilon_{ik}^{v})f_{ik}(\mathbf{r'}-\mathbf{r})+(\epsilon_{ll}^{s}+\epsilon_{ll}^{s'}+\epsilon_{ll}^{v}+
\epsilon_{ll}^{v'})U(|\mathbf{r}-\mathbf{r'}|)\rho_{0}(\mathbf{r})\rho_{0}(\mathbf{r'})d\mathbf{r}d\mathbf{r'}\nonumber\\
\label{elastic1}\eea
It can be averaged directly. However it involves both quadratic and linear terms, they can be grouped and simplified and hence, the elastic part I of the
Lagrangian reduces to,
\beq \textit{L}_{u}=\int\left(\frac{1}{2}c_{ik}^{(2)}\epsilon_{ik}^{s}\epsilon_{ll}^{s}-\mu \rho\epsilon_{ll}^{s}\right)d\mathbf{r}+\int
\left(\frac{1}{2}c_{ik}^{(2)}\epsilon_{ik}^{v}\epsilon_{ll}^{v}-\mu \rho\epsilon_{ll}^{v}\right)d\mathbf{r}\label{lu}\eeq
where $c_{ik}^{(2)}$ is the elastic constant entering through the quadratic term, and is given by
\beq c_{ik}^{(2)}=\frac{1}{V}\int_{V}\frac{\hbar^{2}}{2m}\frac{\partial_{i}\rho_{0}\partial_{k}\rho_{0}}{\rho_{0}}d\mathbf{r} \nonumber\eeq
and
\beq \mu=\frac{\hbar^{2}}{4m}\left(\frac{(\nabla\rho_{0})^{2}}{2\rho_{0}^{2}}-\frac{\nabla^{2}\rho_{0}}{\rho_{0}}\right)
+\int U(|\mathbf{r}-\mathbf{r'}|)\rho_{0}(\mathbf{r'})
d\mathbf{r'}\label{muexpress}\eeq
The chemical potential defined in equation (\ref{muexpress}) is for the usual Gross Pitaevskii equation with long range interaction. $\rho_{0}$ is the ground state density in terms of which the chemical potential is defined.

(4) $\textit{L}_{\tilde{\phi}}$ is the \textbf{hydrodynamical part II}, given as
\bea \textit{L}_{\tilde{\phi}}& = & \hbar\int\rho_{0}(\mathbf{r})(\partial_{t}u_{k}^{ss}\partial_{k}\tilde{\phi}+\partial_{t}u_{k}^{v}\partial_{k}
\tilde{\phi}+\frac{\hbar}{m}\partial_{i}u_{k}^{ss}\partial_{i}\Phi\partial_{k}\tilde{\phi}
+\frac{\hbar}{m}\partial_{i}u_{k}^{v}\partial_{i}\Phi\partial_{k}\tilde{\phi}\nonumber\\
\mbox{} & & -\frac{\hbar}{m}\nabla\phi\cdot\nabla\tilde{\phi}-\frac{\hbar}{2m}(\nabla\tilde{\phi})^{2})d\mathbf{r}
\label{hyd2new1}\eea
Now, above equation (\ref{hyd2new1}) can be re-written as
\bea \textit{L}_{\tilde{\phi}}& = & =-\frac{\hbar^{2}}{2m}\int(2\rho_{0}\mathbf{A}^{s}\cdot\nabla\tilde{\phi}+2\rho_{0}\mathbf{A}^{v}
\cdot\nabla\tilde{\phi}+
\rho_{0}(\nabla\tilde{\phi})^{2})d\mathbf{r}\nonumber\eea
with
\beq \mathbf{A}^{s}=\left(\nabla\phi-(\nabla\phi\cdot\nabla)\mathbf{u}^{ss}-\frac{m}{\hbar}\partial_{t}\mathbf{u}^{ss}\right)\label{aseq}\eeq
and,
\beq \mathbf{A}^{v}=
\left(-(\nabla\phi\cdot\nabla)\mathbf{u}^{v}-\frac{m}{\hbar}\partial_{t}\mathbf{u}^{v}\right)\label{aveq}\eeq
The Euler-Lagrange condition for this part of Lagrangian $\textit{L}_{\tilde{\phi}}$ is
\beq \mathbf{A}^{s}\cdot\nabla\rho_{0}+\mathbf{A}^{v}\cdot\nabla\rho_{0}+\nabla\cdot(\rho_{0}\nabla\tilde{\phi})=0\nonumber\eeq
Solving this equation  for $\tilde{\phi}$ we get $\tilde{\phi}=K_{i}(A_{i}^{s}+A_{i}^{v})$ with $\mathbf{K(r)}$ is a periodic function \cite{Rica2} which satisfies $\nabla_{i}\rho_{0}+\nabla\cdot(\rho_{0}\nabla K_{i})=0$.
%
Above contribution to the Lagrangian can be written in simplified form as \cite{Rica2}
\bea \textit{L}_{\tilde{\phi}}& = & \frac{\hbar^{2}}{2m}\int\rho^{c}_{ij}A_{i}^{s}A_{j}^{s}+\rho^{v}_{ij}A_{i}^{v}A_{j}^{v}d\mathbf{r}\label{hyd2new}\eea
with $\rho^{c}_{ij}$ is the tensor which for symmetric crystal structures is $\rho^{c}_{ij}=\rho\delta_{ij}-\rho_{ij}^{ss}$, defined as
\bea \rho^{c}_{ij}&=& \frac{1}{V}\int_{V}\rho_{0}(\mathbf{r})(\nabla K_{i}\cdot\nabla K_{j})d\mathbf{r}\nonumber\eea
The quantity $\rho^{ss}=\rho$ if the crystal modulation is absent. It is to note that we neglected the last term in equation (\ref{hyd2new}) (proportional to $A_{i}^{s}A_{i}^{v}$).  because we donot want to take into account the vortex crystal effective mass. We only consider the mass density of the supersolid lattice and
the superfluid component in the system. However when this term is included it will
probably give rise to terms with interaction between the two lattices. In present set
of calculations, we assume both the lattices to be independent of each other and hence, amounts to ignoring  a direct coupling between the 
supersolid lattice and the vortex lattice.

(5) $\textit{L}_{\tilde{\rho}}$ is the \textbf{elastic part II}, given by
\bea \textit{L}_{\tilde{\rho}}& = & \frac{\hbar^{2}}{4m}\int\left(\frac{(\nabla\rho_{0})^{2}}{2\rho_{0}^{2}}\tilde{\rho}
-\frac{\partial_{i}\rho_{0}}{\rho_{0}} \partial_{i}\tilde{\rho}
\right)d\mathbf{r}-\frac{1}{2}\int U(|\mathbf{r}-\mathbf{r'}|)(\rho_{0}(\mathbf{r})\tilde{\rho}(\mathbf{r'})\nonumber\\
\mbox{} & & +\tilde{\rho}(\mathbf{r})\rho_{0}(\mathbf{r'})d\mathbf{r}d\mathbf{r'}\label{last1}\\
\mbox{} & & -\frac{\hbar^{2}}{8m}
\int\left(-2(\epsilon_{ik}^{s}+\epsilon_{ik}^{v})\frac{\partial_{i}\rho_{0}\partial_{k}\rho_{0}}{\rho_{0}^{2}}\tilde{\rho}+4(\epsilon_{ik}^{s}+
\epsilon_{ik}^{v})\frac{\partial_{i}\rho_{0}}{\rho_{0}}\partial_{k}\tilde{\rho}+\frac{1}{\rho_{0}}(\nabla\tilde{\rho})^{2}\right)d\mathbf{r}\nonumber\\
\mbox{} & & -\frac{1}{2}\int U(|\mathbf{r}-\mathbf{r'}|)\tilde{\rho}(\mathbf{r})\tilde{\rho}(\mathbf{r'})d\mathbf{r}d\mathbf{r'}\nonumber\\
\mbox{} & & -\frac{1}{2}\int((\epsilon_{ik}^{s}+\epsilon_{ik}^{v})
f_{ik}(\mathbf{r}-\mathbf{r'}))+((\epsilon_{ll}^{s}+\epsilon_{ll'}^{s})+(\epsilon_{ll}^{v}+\epsilon_{ll'}^{v}))U(|\mathbf{r}-\mathbf{r'}|) \nonumber\\
& &(\rho_{0}(\mathbf{r})\tilde{\rho}(\mathbf{r'})+\tilde{\rho}(\mathbf{r})\rho_{0}(\mathbf{r'})
 d\mathbf{r}d\mathbf{r'} \label{elas2}\eea
The terms which are quadratic in the gradients of $\tilde{\rho}$ are the relevant terms because the terms linear in $\tilde{\rho}$ disappears and the action is at minimum when $n=\rho_{0}(\mathbf{r})$ (see equation (\ref{rhoans}). Also, the line (\ref{last1}) in above equation is equal to $-\mu\int\tilde{\rho}
(\mathbf{r})d\mathbf{r}$. Thus, keeping only relevant terms as below:
\bea \frac{\hbar^{2}}{4m}\nabla\cdot\left(\frac{\nabla\tilde{\rho}}{\rho_{0}}\right)-\int U(|\mathbf{r}-\mathbf{r'}|)\tilde{\rho}
(\mathbf{r'})d\mathbf{r'} &=&
\mu(\epsilon_{kk}^{s}+\epsilon_{kk}^{v})\nonumber\\
\mbox{} & & +\frac{\hbar^{2}}{4m}(\epsilon_{ik}^{s}+\epsilon_{ik}^{v})\left(\frac{\partial_{i}\rho_{0}\partial_{k}\rho_{0}}
{\rho_{0}^{2}}-2\frac{\partial_{ik}\rho_{0}}{\rho_{0}}\right)\nonumber\\
\mbox{} & & +\epsilon_{ik}\int (f_{ik}(\mathbf{r}-\mathbf{r'})+2\delta_{ik}U(|\mathbf{r}-\mathbf{r'}|))\rho_{0}(\mathbf{r'})d\mathbf{r'}\nonumber\\
\label{elas3}\eea
The solution of above equation is periodic function $E_{ik}(\mathbf{r})$ \cite{Rica2,homogenization} and of the form $\tilde{\rho}=\epsilon_{ik}E_{ik}(\mathbf{r})$ \cite{Rica3}. Putting in expression
 (\ref{elas3}) and adding the expression (\ref{lu}) we get
\beq \textit{L}_{u}+\textit{L}_{\tilde{\rho}}=-\frac{1}{2}\int(\lambda_{iklm}^{s}\epsilon_{ik}^{s}\epsilon_{lm}^{s}
+\lambda_{iklm}^{v}\epsilon_{ik}^{v}\epsilon_{lm}^{v})d\mathbf{r}\label{mod1}\eeq
where $\lambda_{iklm}$ is given by
\bea \lambda_{iklm}^{s}&=&-\frac{1}{V}\int_{V}\frac{\hbar^{2}}{2m}\frac{\partial_{i}\rho_{0}\partial_{k}\rho_{0}}
{\rho_{0}}\delta_{lm}d\mathbf{r}-\frac{1}{V}\int_{V}\mu(\delta_{ik}E_{lm}^{s}(\mathbf{r})+\delta_{lm}E_{ik}^{s}(\mathbf{r}))d\mathbf{r}\nonumber\\
\mbox{} & & -\frac{1}{V}\int_{V}d\mathbf{r}\left(\frac{\hbar^{2}}{4m}\frac{1}{\rho_{0}}(\nabla E_{ik}^{s})\cdot(\nabla E_{lm}^{s})+\int U(|\mathbf{r}-\mathbf{r'}|)
E_{ik}^{s}(\mathbf{r})E_{lm}^{s}(\mathbf{r'})d\mathbf{r'}\right)\nonumber\\
\label{lambdas}\eea
and,
\bea \lambda_{iklm}^{v}&=&-\frac{1}{V}\int_{V}\frac{\hbar^{2}}{2m}\frac{\partial_{i}\rho_{0}\partial_{k}\rho_{0}}
{\rho_{0}}\delta_{lm}d\mathbf{r}-\frac{1}{V}\int_{V}\mu(\delta_{ik}E_{lm}^{v}(\mathbf{r})+\delta_{lm}E_{ik}^{v}(\mathbf{r}))d\mathbf{r}\nonumber\\
\mbox{} & & -\frac{1}{V}\int_{V}d\mathbf{r}\left(\frac{\hbar^{2}}{4m}\frac{1}{\rho_{0}}(\nabla E_{ik}^{v})\cdot(\nabla E_{lm}^{v})+\int U(|\mathbf{r}-\mathbf{r'}|)
E_{ik}^{v}(\mathbf{r})E_{lm}^{v}(\mathbf{r'})d\mathbf{r'}\right)\nonumber\\
\label{lambdav}\eea
The expression $\frac{1}{2}\lambda_{iklm}\epsilon_{ik}\epsilon_{lm}$ is the expression for the elastic energy density of a solid, and $\lambda_{iklm}$ is the
elastic modulus tensor \cite{landau}.

Hence, we can write the effective Lagrangian for the long wave perturbations of displacement of both lattices, of average density and of the phase as the sum of
various contribution mentioned above.
\bea \textit{L}_{eff}&=&\int d\mathbf{r} [-\hbar \rho\frac{\partial\phi}{\partial t}-\frac{\hbar^{2}}{2m}\left[\rho(\nabla\phi)^{2}-(\rho\delta_{ik}-\rho_{ik}^{ss})
\left(\nabla\phi-\frac{m}{\hbar}\frac{D\mathbf{u}^{ss}}{Dt}\right)_{i}
\left(\nabla\phi-\frac{m}{\hbar}\frac{D\mathbf{u}^{ss}}{Dt}\right)_{k}\right]\nonumber\\
& & \mbox{} -\mathcal{E}(\rho)-\frac{1}{2}\lambda_{iklm}^{s}\epsilon_{ik}^{s}\epsilon_{lm}^{s}-\frac{1}{2}\lambda_{iklm}^{v}\epsilon_{ik}^{v}
\epsilon_{lm}^{v}-m\rho\mathbf{v}_{s}\cdot(\Omega\times\mathbf{r})]\label{leff}\eea
where
\beq \frac{D\mathbf{u}^{ss}}{Dt}=\frac{\partial\mathbf{u}^{ss}}{\partial t}+\frac{\hbar}{m}\nabla\phi\cdot\nabla\mathbf{u}^{ss}\label{dussdt}\eeq
The above Lagrangian can also be written as
\beq \textit{L}_{eff}=\int d\mathbf{r} \left[-\hbar \rho \frac{\partial \phi}{\partial t}-\mathcal{E}\right] \nonumber \eeq
where
\bea \mathcal{E}=\mathcal{E}_{in}(\rho)+\mathcal{E}_{ph}(\phi)+\mathcal{E}_{el}^{ss}(\nabla\mathbf{u}^{ss})+\mathcal{E}_{el}^{v}(\nabla\mathbf{u}^{v})
\label{endens2}\eea
with
\bea \mathcal{E}_{in}(\rho)&=& \frac{\hbar^{2}}{2m}\frac{(\nabla\rho)^{2}}{4\rho}+\rho\int U(|\mathbf{r}-\mathbf{r}'|)\rho(\mathbf{r}')d\mathbf{r}'=\mu\rho
\label{endens21}\\
\mathcal{E}_{ph}(\phi)&=&\frac{\hbar^{2}}{2m}\left[\rho(\nabla\phi)^{2}-(\rho\delta_{ik}-\rho_{ik}^{ss})
\left(\nabla\phi-\frac{m}{\hbar}\frac{D\mathbf{u}^{ss}}{Dt}\right)_{i}
\left(\nabla\phi-\frac{m}{\hbar}\frac{D\mathbf{u}^{ss}}{Dt}\right)_{k}\right]\label{endens22}\\
\mbox{} & & +m\rho\mathbf{v}_{s}\cdot(\Omega\times\mathbf{r})\nonumber\\
\mathcal{E}_{el}^{ss}(\nabla\mathbf{u}^{ss})&=& \frac{1}{2}\lambda_{iklm}^{s}\epsilon_{ik}^{s}\epsilon_{lm}^{s}\label{endens23}\\
\mathcal{E}_{el}^{v}(\nabla\mathbf{u}^{v})&=& \frac{1}{2}\lambda_{iklm}^{v}\epsilon_{ik}^{v}\epsilon_{lm}^{v}\label{endens24}\eea

For the usual superfluid, the Gross-Pitaevskii equation is recovered by above Lagrangian. When there is no crystal lattice (either vortex lattice or supersolid crystal lattice), the equations (\ref{endens23}) and (\ref{endens24}) have no contribution and similarly, the second and third terms in equation (\ref{endens22}) vanishes alongwith the long range interaction term in equation (\ref{endens21}), hence recovering the Lagrangian for cold atomic superfluids.

It can be clearly seen from the above equation (\ref{leff}) that the crystal lattice may have a different velocity than the superfluid component,
with the velocity difference proportional to $\left(\nabla\phi-\frac{\hbar}{m}\frac{D\mathbf{u}^{ss}}{Dt}\right)$,
with $\mathbf{u}^{ss}$ as the displacement field of crystal lattice due to density modulations in superfluid. Hence the third term in equation
(\ref{leff}) gives the product of the mass density of the supersolid lattice and the square of the supersolid lattice
velocity.

Here $\rho_{ik}^{ss}$ is the superfluid density tensor \cite{Paananen}, which is in general a symmetric matrix.
In our further calculations, we express that the superfluid density is a function of local number density $\rho$ and for isotropic symmetry of lattice,
it is given by $\rho_{ik}^{ss}=\rho^{ss}(\rho)\delta_{ik}$.

It may be noted from the structure of the proposed Lagrangian that we donot take into account coupling of the two lattices with
displacement fields $\mathbf{u}^{ss},\mathbf{u}^{v}$ in the lowest order
expansion and thus the elastic deformations of the two lattices do not interact with each
other directly.

\subsection{Hydrodynamic equations of motion for rotating supersolid}
Here we provide the detailed derivations of the hydrodynamic equation of a rotating supersolid that appears in the main paper
through the extremization of the hydrodynamic Lagrangian. 
The dynamical equations are derived by variation of action $S=\int\mathcal{L}dt$ taken as a functional of $\rho$, $\phi$, $\mathbf{u}^{ss}$ and $\mathbf{u}^{v}$.
This yields a set of coupled of partial differential equations for those fields.
The action to be extremised is $S=\int\textit{L}dt $, gives the condition
\bea \delta\int\textit{L}dt =0\nonumber\eea
where, $\textit{L}=\textit{L}(\rho,\phi,\nabla\mathbf{u}^{ss},\nabla\mathbf{u}^{v})$, which implies
\bea \frac{\partial\textit{L}}{\partial\rho}d\rho+\frac{\partial\textit{L}}{\partial\phi}d\phi+\frac{\partial\textit{L}}
{\partial(\nabla\mathbf{u}^{ss})}d(\nabla
\mathbf{u}^{ss})+\frac{\partial\textit{L}}{\partial(\nabla\mathbf{u}^{v})}d(\nabla\mathbf{u}^{v})=0\label{fulllag}\eea

We calculate,
\bea \frac{\partial\textit{L}}{\partial\rho}&=&-\hbar\frac{\partial\phi}{\partial t}-\left[\frac{\nabla^{2}\rho}{2\rho}-\frac{(\nabla\rho)^{2}}{4\rho^{2}}\right]
\frac{\hbar^{2}}{2m}-\int U(|\mathbf{r}-\mathbf{r}'|)\rho(\mathbf{r}')d\mathbf{r}'-\frac{\hbar^{2}}{2m}(\nabla\phi)^{2}\nonumber\\
\mbox{} & & +m\mathbf{v_{s}}\cdot(\Omega\times\mathbf{r})=0\nonumber\eea
Taking gradient on both sides,
\bea \frac{\partial\mathbf{v_{s}}}{\partial t}+\nabla\left(\frac{1}{2}\mathbf{v_{s}}^{2}\right)=-\nabla\left[\frac{T+V+g\rho+\int U(|\mathbf{r}-\mathbf{r}'|)\rho(\mathbf{r}')d\mathbf{r}'}{m}-\mathbf{v_{s}}\cdot(\Omega\times\mathbf{r})\right]\label{roteuler}\eea
where $\mathbf{v_{s}}$ is the superfluid velocity defined as $\mathbf{v_{s}}=\frac{\hbar}{m}\nabla\phi$.\\
Next,
\bea \frac{\partial\textit{L}}{\partial\phi}=-\hbar\frac{\partial\rho}{\partial t}-\frac{\partial}{\partial\phi}[\mathcal{E}_{ph}(\phi)]=0\label{conti1}\eea
We determine the second term in above equation separately,
\bea \frac{\partial}{\partial\phi}[\mathcal{E}_{ph}(\phi)]&=&\frac{\partial}{\partial x_{i}}\left[\rho\frac{\hbar^{2}}{m}\nabla\phi\right] -\frac{\partial}{\partial\phi}\left[(\rho\delta_{ik}-\rho_{ik}^{ss})\left(\nabla\phi-\frac{m}{\hbar}\dot{\mathbf{u_{s}}}\right)_{i}
\left(\nabla\phi-\hbar\frac{m}{\hbar}\dot{\mathbf{u_{s}}}\right)_{k}\right]\nonumber\\
\mbox{} & &-\frac{\partial}{\partial x_{i}}[\rho\cdot(\Omega\times\mathbf{r})]\nonumber\\
&=& \frac{\partial}{\partial x_{i}}\left[\rho\frac{\hbar^{2}}{m}\nabla\phi\right] -\hbar\frac{\partial}{\partial x_{i}}\left[(\rho\delta_{ik}-\rho_{ik}^{ss})\left(\frac{\hbar}{m}\partial_{i}\phi-\dot{\mathbf{u_{s}}}\right)_{i}\right]\nonumber\\
& & \mbox{} -\hbar\frac{\partial}{\partial x_{i}}[\rho\cdot(\Omega\times\mathbf{r})]\eea
Putting the above expression for $\frac{\partial}{\partial\phi}[\mathcal{E}_{ph}(\phi)]$ in equation (\ref{conti1}), we get
\bea \frac{\partial\rho}{\partial t}+\frac{\partial}{\partial x_{i}}\left[\rho\frac{\hbar}{m}\nabla\phi\right]+\frac{\partial}{\partial x_{k}}\left[(\rho-\rho^{ss})(\delta_{ik}
-\partial_{k}u_{i}^{ss})\left(\dot{u}_{i}^{ss}-\frac{\hbar}{m}\partial_{i}\phi\right)\right] -\frac{\partial}{\partial x_{i}}[\rho\cdot(\Omega\times\mathbf{r})]&=& 0\nonumber\\
\label{conti2}\eea

Equations (\ref{roteuler}) and (\ref{conti2}) are the modified Euler and Continuity equation for a condensate rotating at an angular frequency $\Omega$. The energies in the laboratory and rotating frame are related by $ \mathcal{E}_{R} = \mathcal{E} - \Omega\cdot \mathbf{r} \times \mathbf{p} $. The transformation to rotating frame of reference introduces the $\mathbf{v_{s}}\cdot\Omega\times\mathbf{r}$ term and $\rho\cdot\Omega\times\mathbf{r}$ term in equation (\ref{roteuler}) and (\ref{conti2}) respectively \cite{labrotref}.  Here, $\mathbf{v_{s}}$ is the superfluid velocity in the lab (inertial) frame of reference. In the rotating frame of reference, these equations along with the equations of elastic response of the system due to the supersolid lattice and the vortex lattice are given as
\bea \frac{\partial\rho}{\partial t}+\nabla\cdot\left(\rho\frac{\hbar}{m}\nabla\phi\right)+\frac{\partial}{\partial x_{k}}\left[(\rho-\rho^{ss})(\delta_{ik}
-\partial_{k}u_{i}^{ss})\left(\dot{u}_{i}^{ss}-\frac{\hbar}{m}\partial_{i}\phi\right)\right]&=& 0\label{aphyd1}\eea
and,
\bea m\left(\frac{\partial \mathbf{v_{s}}}{\partial t}+2\mathbf{\Omega}\times\mathbf{v}_{s}\right)=-\frac{\nabla P'}{\rho}\label{phirot3}\eea
Here, $\rho_{ik}^{ss}$ is the superfluid density tensor \cite{Paananen} which assumes the form $\rho_{ik}^{ss}=\rho^{ss}\delta_{ik}$ for isotropic symmetry of the system and,
\beq P'=\rho\left(T+\int U(|\mathbf{r}-\mathbf{r}'|)\rho(\mathbf{r}')d\mathbf{r}'\right)\label{pprimeexp}\eeq
In the equation (\ref{phirot3}), we have kept only the linearized terms and the nonlinear terms with higher orders of derivatives have been dropped. The neglected
terms in equation (\ref{phirot3}) are given below.
\bea \frac{\partial}{\partial\rho}(\rho-\rho_{ss})\left(\nabla\phi-\frac{m}{\hbar}\frac{D\mathbf{u}^{ss}}{Dt}\right)^{2}
&=&\left(\frac{(\rho^{2}-1)\delta\rho}{\rho}-\nabla\cdot\mathbf{u}^{ss}\right)\left(\nabla\phi-\frac{m}{\hbar}\frac{D\mathbf{u}^{ss}}{Dt}\right)^{2}\nonumber\\
&=&\left(\frac{(\rho^{2}-1)\delta\rho}{\rho}-\nabla\cdot\mathbf{u}^{ss}\right)((\nabla\phi)^{2}+\left(\frac{m}{\hbar}\right)^{2}\left(\frac{D\mathbf{u}^{ss}}{Dt}\right)^{2}
\nonumber\\
\mbox{} & & -2\frac{m}{\hbar}\nabla\phi\cdot\frac{D\mathbf{u}^{ss}}{Dt})\nonumber\eea
When averaged over a vortex lattice cell, equation (\ref{phirot3}) can be written as
\bea m\left(\frac{\partial \mathbf{v_{s}}}{\partial t}+\tilde{\mathbf{\omega}}\times\mathbf{v}_{L}\right)=-\frac{\nabla P'}{\rho}\label{aphyd2}\eea
with $\mathbf{v_{s}}$ as the averaged velocity and $\mathbf{\tilde{\omega}}=2\mathbf{\Omega}+\nabla\times\mathbf{v}_{s}$ as the averaged vorticity \cite{Sonin}. The
velocity of the vortex is given by $\mathbf{v}_{L}$ and it is equal to time derivative of the displacement vector of the vortex lattice
$\dot{\mathbf{u}}^{v}$.

The force $\mathbf{f}$ acting per unit volume of the fluid moving with
velocity $\mathbf{v}_{s}$ is
\bea \mathbf{f}_{el}^{v}=-\rho\mathbf{\tilde{\omega}}\times(\mathbf{v}_{L}-\mathbf{v}_{s})\label{force}\eea
and it should be connected with a variation of energy due to vortex displacements. Thus,
\bea f_{el,i}^{v}&=&-\frac{\delta\mathcal{E}_{el}^{v}}{\delta u_{i}^{v}}=-\frac{\partial}{\partial x_{k}}\left(\frac{\delta\mathcal{E}_{el}^{v}}
{\delta(\partial u_{k}^{v}/\partial x_{i})}\right)\nonumber\\
&=& -[(\lambda^{v}+\mu_{s}^{v})\nabla(\nabla\cdot\mathbf{u}^{v})+\mu_{s}^{v}\nabla^{2}
\mathbf{u}^{v}]\label{vorlatticeelas} \eea

Hence using equations (\ref{force}) and (\ref{vorlatticeelas}) we get
\bea \rho 2\Omega[\hat{z}\times(\mathbf{v}_{L}-\mathbf{v}_{s})]=\frac{(\lambda^{v}+\mu_{s}^{v})\nabla(\nabla\cdot\mathbf{u}^{v})+\mu_{s}^{v}\nabla^{2}
\mathbf{u}^{v}}{m}  \label{aphyd3}\eea
where $\lambda^{v}=K^{v}-\frac{2}{3}\mu_{s}^{v}$ is the Lam\'{e} coefficient, and $K^{v}$ and $\mu_{s}^{v}$ are the
compressibility and shear modulus of the vortex lattice. Equation (\ref{hyd3}) is the equation of motion of the system due to the elastic response of the vortex lattice.

Next we determine the equation of motion due to the elastic response of the supersolid crystal lattice by considering the $4$th term in equation
(\ref{fulllag}). Finally,
\bea \frac{\partial\mathcal{L}}{\partial\mathbf{u_{s}}}&=& m\frac{\partial}{\partial t}\left[(\rho\delta_{ik}-\rho_{ik}^{ss})\left(\dot{u_{i}^{ss}}-\frac{\hbar}{m}\partial_{k}\phi\right)\right]+\hbar\frac{\partial}{\partial x_{k}}\left[(\rho\delta_{ik}-\rho_{ik}^{ss})\left(\dot{u_{i}^{ss}} -\frac{\hbar}{m}\partial_{i}\phi\right)\partial_{k}\phi\right]\nonumber\\
\mbox{} & &+\frac{\partial}{\partial x_{k}}(\lambda_{iklm}^{ss}\epsilon_{lm}^{ss})=0\label{aphyd41}\eea
Thus, 
\bea m\frac{\partial}{\partial t}\left[(\rho-\rho^{ss})\left(\dot{u}_{i}^{ss}-\frac{\hbar}{m}\partial_{i}\phi\right)\right]+
\hbar\frac{\partial}{\partial x_{k}}\left[(\rho-\rho^{ss})\left(\dot{u}_{i}^{ss} -\frac{\hbar}{m}\partial_{i}\phi\right)\partial_{k}\phi\right]\nonumber\\
\mbox{} +\frac{\partial}{\partial x_{k}}(\lambda_{iklm}^{ss}\epsilon_{lm}^{ss})=0\label{hyd41}\eea
Here too, when the lattice is assumed to be isotropic then above equation (\ref{hyd41}) can be written as
\bea m\frac{\partial}{\partial t}\left[(\rho-\rho^{ss})\left(\dot{u}_{i}^{ss}-\frac{\hbar}{m}\partial_{i}\phi\right)\right]
+\hbar\frac{\partial}{\partial x_{k}}\left[(\rho-\rho^{ss})\left(\dot{u}_{i}^{ss} -\frac{\hbar}{m}\partial_{i}\phi\right)\partial_{k}\phi\right]\nonumber\\
\mbox{} +(\lambda^{ss}+\mu_{s}^{ss})\partial_{ik}u_{k}^{ss}+\mu_{s}^{ss}\nabla^{2}u_{i}^{ss} & =& 0\nonumber\\
\label{aphyd4}\eea
where $\lambda^{ss}=K^{ss}-\frac{2}{3}\mu_{s}^{ss}$ is the second Lame coefficient, and $K^{ss}$ and $\mu_{s}^{ss}$
are the compressibility and shear modulus of the solid \cite{landau}.

Thus, equations (\ref{aphyd1}), (\ref{aphyd2}), (\ref{aphyd3}) and (\ref{aphyd4}) are the equations of motion for a rotating supersolid with elastic properties of both
vortex lattice and supersolid crystal lattice taken into account.

\bea \frac{\partial\rho}{\partial t}+\nabla\cdot\left(\rho\frac{\hbar}{m}\nabla\phi\right)+\frac{\partial}{\partial x_{k}}
\left[(\rho-\rho^{ss})(\delta_{ik}-\partial_{k}
u_{i}^{ss})\left(\dot{u}_{i}^{ss}-\frac{\hbar}{m}\partial_{i}\phi\right)\right]&=& 0\label{hydmixeq1}\eea
\bea m\left(\frac{\partial\mathbf{v_{s}}}{\partial t}+2\mathbf{\Omega}\times\mathbf{v}_{L}\right)=-\frac{\nabla P'}{\rho}\label{hydmixeq2}\eea
\bea \rho 2\Omega[\hat{z}\times(\mathbf{v}_{L}-\mathbf{v}_{s})]=\frac{(\lambda^{v}+\mu_{s}^{v})\nabla(\nabla\cdot\mathbf{u}^{v})
+\mu_{s}^{v}\nabla^{2}\mathbf{u}^{v}}{m} \label{hydmixeq3}\eea
\bea m\frac{\partial}{\partial t}\left[(\rho-\rho^{ss})\left(\dot{u}_{i}^{ss}-\frac{\hbar}{m}\partial_{i}\phi\right)\right]+
\hbar\frac{\partial}{\partial x_{k}}\left[(\rho-\rho^{ss})\left(\dot{u}_{i}^{ss} -\frac{\hbar}{m}\partial_{i}\phi\right)\partial_{k}\phi\right]\nonumber\\
\mbox{} -(\lambda^{ss}+\mu_{s}^{ss})\partial_{ik}u_{k}^{ss}+\mu_{s}^{ss}\nabla^{2}u_{i}^{ss} &=& 0\nonumber\\
\label{hydmixeq4}\eea
These set of four equations form the hydrodynamic equations of motion for a rotating supersolid system.



\begin{thebibliography}{99}
\bibitem{Mosechan} E. Kim and M. H. W Chan, Science {\bf 305}, 1941 (2004)
\bibitem{Reppy} J. D. Reppy, Phys. Rev. Lett. {\bf 104} 255301 (2010).
\bibitem{Mosechan1}D. Y. Kim and M. H. W. Chan, Phys. Rev. Lett. {\bf 109}, 155301 (2012).
\bibitem{PO}O. Penrose and L. Onsager, Phys. Rev. {\bf 104}, 576 (1956).
\bibitem{Andreev} A. F. Andreev and I. M. Lifshitz, JETP {\bf 29}, 1107 (1969).
\bibitem{Legget}A. J. Leggett, Phys. Rev. Lett.  {\bf 25}, 1543 (1970).
\bibitem{ODLRO} H. Matsuda and T. Tsuneto, Prog. Theor. Phys. Suppl. {\bf 46}, 411 (1970).
\bibitem{Anderson}P. W. Anderson, W. F. Brinkman, D. A. Huse {\bf 310}, 1164 (2005).
\bibitem{review}N. Prokof'ev, Adv. Phys. {\bf 56}, 381 (2007); N. Prokof'ev and M. Boninsegni, Rev. Mod. Phys. {\bf 84}, 759 (2012). 
\bibitem{santos} L. Santos, G. V. Shlyapnikov, and M. Lewenstein, Phys. Rev. Lett. {\bf 90}, 250403 (2003)
\bibitem{Pomeau}Y. Pomeau and S. Rica, Phys. Rev. Lett. {\bf 72}, 2426 (1994).
\bibitem{Nozieres}P. Nozieres, J. Low Temp. Phys {\bf 137}, 45 (2004).
\bibitem{dipole1}T. Lahaye {\it et al.}, Nature (London) {\bf 448}, 672 (2007).
\bibitem{Rydberg}R. Heidermann {\it et al.}, Phys. Rev. Lett. {\bf 100}, 033601 ( 2008). 
\bibitem{polar1}K.-K. Ni {\it et al.}, Science 322, 231 (2008).
\bibitem{Grimm}K. Aikawa {\it et al.}, Phys. Rev. Lett. {\bf 108}, 210401 (2012)
\bibitem{Lev}M. Lu {\it et al.}, Phys. Rev. Lett. {\bf 108}, 215301 (2012).
\bibitem{He}L. He and W. Hofstetter, Phys. Rev. A {\bf 83}, 053629 (2011).
\bibitem{Jain}P. Jain, F. Cinti and M. Boninsegni, Phys. Rev. B {\bf 84}, 014534 (2011).
\bibitem{Rejish}N. Henkel, R. Nath and T. Pohl, Phys. Rev. Lett  {\bf  104} , 195302 (2010).
\bibitem{Cinti}F. Cinti {\it et al.}, Phys. Rev. Lett. {\bf 105}, 135301 (2010).
\bibitem{Li}X. Li {\it et al.}, Phys. Rev. Lett. {\bf 83}, 012602(R) (2011). 
\bibitem{RotonEx}R. Motti {\it et al.} Science {\bf 336}, 1570 (2012).
\bibitem{expt2}K. Baumann, C. Guerlin, F. Brennecke, and T. Esslinger, Nature (London)  {\bf 464}, 1301 (2010).
\bibitem{rotation}K.W. Madison, F. Chevy, W.Wohlleben, and J. Dalibard, Phys. Rev. Lett. {\bf 84}, 806 (2000); J. R. Abo-Sheer et al., Science {\bf 292}, 476 (2001); P. Engels, I. Coddington, P. C. Haljan, and E. A. Cornell, Phys Rev. Lett. {\bf 89}, 100403 (2002).
\bibitem{rotlat}S. Tung, V. Schweikhard, and E. A. Cornell, Phys. Rev. Lett. {\bf 97}, 240402 (2006); R. A. Williams, S. Al Assam, and C. J. Foot,
{\it ibid.}  {\bf 104}, 050404 (2010).
\bibitem{synthetic} Y-J. Lin et al., Nature , {\bf 462}, 628 (2009); Y. J. Lin, R. L. Compton, A. R. Perry, W. D. Phillips, J. V. Porto, and I. B. Spielman, Phys. Rev. Lett. {\bf 102}, 130401 (2009); J. Dalibard et.al, arXiv 1008.5378
\bibitem{Mason}P. Mason, C. Josserand and S. Rica, Phys. Rev. Lett. {\bf 109}, 045301 (2012). 
\bibitem{ssvortex}N. Henkel {\it et al.}, Phys. Rev. Lett. {\bf 108},  265301 (2012).
\bibitem{Gross}E. P. Gross, Phys. Rev., {\bf 106}, 161 (1957); E. P. Gross, Ann. Phys. (N.Y.), {\bf 4}, 57 (1958)
\bibitem{Coddington}I. Coddington, P. Engels, V. Schweikhard, and E. A. Cornell, Phys. Rev. Lett. {\bf 91}, 100402 (2003).
\bibitem{tkachenko} V. K. Tkachenko, Zh. Eksp. Teor. Fiz. {\bf 49}, 1875 (1965) [Sov. Phys. JETP 22, {\bf 1282} (1966)]; Zh. Eksp. Teor. Fiz. {\bf 50}, 1573 (1966) [Sov. Phys. JETP {\bf 23}, 1049 (1966)]; Zh. Eksp. Teor. Fiz. {\bf 56}, 1763 (1969) [Sov. Phys. JETP
{\bf 29}, 245 (1969)]
\bibitem{mizushima} T. Mizushima, Y. Kawaguchi, K. Machida, T. Ohmi, T. Isoshima, and M. M. Salomaa, Phy. Rev. Lett. {\bf 92}, 060407 (2004)
\bibitem{Baym3} G. Baym, E. Chandler, J. of Low Temp. Physics {\bf 50}, 57 (1983).
\bibitem{Sonin1} E. B. Sonin, Rev. Mod. Phys. {\bf 59}, 87 (1987).
\bibitem{Baym}G. Baym, Phys. Rev. Lett, {\bf 91}, 110402 (2003).
\bibitem{Baym1}G. Baym, Phys. Rev. A., { \bf 69}, 043618 (2004). 
\bibitem{Sonin}E. B. Sonin, Phys. Rev. A, {\bf 71}, 011603(R), (2005).
\bibitem{Fetterreview}A. L. Fetter,  Rev. Mod. Phys. , {\bf 81}, 647 (2009).
\bibitem{boninsegni} S. Saccani, S. Moroni, and M. Boninsegni, Phys. Rev. Lett. {\bf 108}, 175301 (2012)
\bibitem{newref} N Goldman, G Juzeliunas, P Ohberg and I B Spielman, Rep. Prog. Phys. {\bf 77}, 126401 (2014)
\bibitem{highrot}V. Schweikhard {\it et al.}, Phys. Rev. Lett. {\bf 92}, 040404 (2004). 
\bibitem{Book1} C. J. Pethick and H. Smith, {\it Bose-Einstein Condensation in Dilute Gases}, ( Cambridge University Press, New York, 2008), Chapter $6$ and Chapter $9$.
\bibitem{Book2}L. Pitaevskii and S. Stringari, {Bose-EInstein Condensation}, (Claredon Press, Oxford 2003), Chapter $5$ amd Chapter $14$.   
\bibitem{Rica2}C. Josserand, Y Pomeau and S. Rica, Eur. Phys. J. Special Topics {\bf 146}, 47 (2007).
\bibitem{newrica} C. Josserand, Y Pomeau and S. Rica, Phys. Rev. Lett. {\bf 98}, 195301 (2007)
\bibitem{Rica3}G. During {\it et al.}, {\it Lecture Notes of the 4th Warshaw school on statistical physics}, cond-mat/arXiv:1110.1323. 
\bibitem{Paananen} T. Paananen, J. Phys. B: At. Mol. Opt. Phys. {\bf 42}, 165304 (2009).
\bibitem{isotropicss} W.  M. Saslow and S. Jolad, Physical Review B {\bf 73}, 092505 (2006); 
N. Sepulveda, C. Josserand, and S. Rica,  Eur. Phys. J. B {\bf 78}, 439 (2010)
\bibitem{supp1} See Appendix A for the
derivation of effective lagrangian  for rotating supersolid in Sec. I; and hydrodynamic equations of motion for rotating supersolid in Sec. II
\bibitem{landau} L. D. Landau and E. M. Lifshitz, 1965, \textit{Theory of Elasticity}


\bibitem{homogenization} A. Bensoussan, J.L. Lions, G. Papanicolaou, \textit{Asymptotic Aanalysis in Periodic Structures} (North-
Holland, Amsterdam, 1978); E. S\'{a}nchez-Palencia, \textit{Non-Homogeneous Media and Vibration Theory}, Lecture Notes Phys. {\bf 127} (Springer-Verlag, Berlin, 1980)
\bibitem{labrotref} James A. Fay, \textit{Introduction to fluid mechanics} (MIT Press, 1994).

%


\end{thebibliography}
\end{document}